\documentclass[10pt,letterpaper]{article}
\usepackage[top=0.85in,left=2.75in,footskip=0.75in]{geometry}

\usepackage[ruled,scleft,nofillcomment]{algorithm2e}
\usepackage{amsmath,amssymb}

\usepackage{textcomp,marvosym}
\usepackage{subfigure}
\usepackage{enumerate}
\usepackage{microtype}
\DisableLigatures[f]{encoding = *, family = * }
\usepackage{lastpage,fancyhdr,graphicx}
\usepackage{epstopdf}

\usepackage{amsmath,amssymb}

\usepackage{changepage}

\usepackage[utf8x]{inputenc}

\usepackage{textcomp,marvosym}

\usepackage{cite}

\usepackage{nameref,hyperref}

\usepackage{microtype}
\DisableLigatures[f]{encoding = *, family = * }

\usepackage[table]{xcolor}

\usepackage{array}

\newcolumntype{+}{!{\vrule width 2pt}}

\newlength\savedwidth

\raggedright
\usepackage[aboveskip=1pt,labelfont=bf,labelsep=period,justification=raggedright,singlelinecheck=off]{caption}

\makeatletter
\renewcommand{\@biblabel}[1]{\quad#1.}
\makeatother

\usepackage{lastpage,fancyhdr,graphicx}
\usepackage{epstopdf}
\fancyheadoffset[L]{2.25in}
\fancyfootoffset[L]{2.25in}
\usepackage{amsthm}
\newtheorem{Definition}{Definition}

\newtheoremstyle{note}
  {3pt}
  {3pt}
  {\slshape}
  {\parindent}
  {\slshape }
  {:}
  {.5em}
  {}%
\theoremstyle{note}
\newtheorem*{example}{Example}

\begin{document}
\vspace*{0.2in}

\begin{flushleft}
{\Large
\textbf\newline{Efficient processing of raster and vector data } 
}
\newline

Fernando Silva-Coira \textsuperscript{1*},
José R. Paramá\textsuperscript{1},
Susana Ladra\textsuperscript{1},
Juan R. López\textsuperscript{1},
 Gilberto Guti\'errez \textsuperscript{2}
\\
\bigskip
\textbf{1} Universidade da Coru\~na,  Centro de investigación CITIC,
Facultade de Inform\'atica,  Campus de Elvi\~na, s/n, A Coru\~na, Spain
\\
\textbf{2} Universidad del B\'io-B\'io,  DCCTI, Chillán, Chile
\bigskip

* fernando.silva@udc.es (FS)
\end{flushleft}

\section*{Abstract}In this work, we propose a framework to store and manage spatial data, which includes new efficient algorithms to perform operations accepting as input a raster dataset and a vector dataset.
More concretely, we present algorithms for solving a spatial join between a raster and a vector dataset imposing a restriction on the values of the cells of the raster; and  an  algorithm for retrieving \textit{K} objects of a vector dataset that overlap  cells  of a raster dataset, such that the $K$ objects are those overlapping the highest (or lowest) cell values among all objects.  
The raster data is stored using a compact data structure, which can directly manipulate compressed data  without the need for prior decompression. This leads to better running times and lower memory consumption. 
In our experimental evaluation comparing our solution to other baselines, we obtain the best space/time trade-offs.

\section*{Introduction}
\label{intro}

When dealing with spatial data, depending on the particular characteristics of the type of  information, it may be more appropriate to represent that information (at the logical level) using either a raster or a vector data model \cite{Couclelis92}.
The advance of the digital society is providing a continuous growth of the amount of available vector data, but  the appearance of cheap devices equipped with GPS, like smartphones, is responsible for a big data explosion, mainly of trajectories of moving objects. The same phenomenon can be found in raster datasets, where the advances in hardware are responsible for an important increment of the size and the amount of available data. Only taking into account the  images acquired by satellites,  several terabytes of data are generated each day \cite{bb78851}, and it has been estimated that the archived amount of raster data    will soon reach the  zettabyte scale \cite{quartulli2013review}. 

This big increase in the variety, richness, and amount of spatial data has also led to new information demands. Nowadays, many application areas require the combination of data stored in different formats \cite{grumbach2000manipulating} to run complex analysis. Obviously, combining different data models becomes more difficult when dealing with large amounts of data.

Although there is a large body of research regarding the size, the analysis, and the heterogeneity of data,  
in the case of spatial data, in most cases, that research is focused either on the vector model or on the raster model separately. The two models are rarely handled together.
For instance, the usual solution for queries that involve (together) raster and vector datasets  is to transform the vector dataset into a raster dataset, and then to use a raster algorithm to solve the query. This is the solution for the zonal statistics operation of Map Algebra in, at least, ArcGIS and GRASS \cite{zonalS,zonalSG}.

However, some previous research has addressed the problem using  a joint approach. In  \cite{grumbach2000manipulating},  a single data model and language is proposed to represent and query both vector and raster data at the logical level. Even a  \emph{Join} operator is suggested, which  allows combining, transparently and interchangeably, vector datasets, raster datasets, or both. As an example, the authors propose the query ``return the coordinates of the trajectory of an aircraft when it was over a ground with altitude over 1,000''. Unfortunately, no implementation details are given.

Other previous contributions deal with the implementation of query operators that are explicitly defined for querying datasets in different formats \cite{corral1999algorithms,TPRtrees,RodriguezBrisaboa17,Eldawy:2017}.  Some of them tackled the Join, or a close query, but in this case, these works suffer from   limitations (data structures not functional enough, too restrictive join operations, size problems) that will be explained more in detail in the next section.\\

On the other hand,  compression has been used traditionally with the aim of just reducing the size of the datasets in disk and during network transmissions. However, it has recently begun to be used as a way to obtain improvements in other dimensions, such as processing time or scalability \cite{plattner2012memory}. In the last few years, several authors \cite{k2ones,SSDBM16,k2rasterIS,pinto2017improved} have proposed the use of modern compact data structures 	\cite{Nav16} to represent raster datasets. Compact data structures use compression to reduce the size of the stored dataset, but with the novelty that the compressed data structure can be managed directly in compressed form, even in main memory. By saving main memory, we obtain a more scalable system, but at the same time, we take advantage of a better usage of the memory hierarchy, and thus obtain better running times. This strategy is sometimes called ``in-memory'' data management  \cite{plattner2013course}. In addition, many compact data structures are equipped with an index that, in the same compressed space, speeds up the queries. This feature is known as ``self-indexation''. One example of these compact data structures designed for raster data, and the one achieving the best space/time trade-offs \cite{pinto2017improved}, is the $k^2$-raster \cite{k2rasterIS}, which will be used in this work, thus extending its functionality. \\

In this work, we propose to use a new framework to  store and manage raster and vector datasets.  The vector dataset is stored and indexed in a traditional way, using an R-tree \cite{Guttman:1984:RDI:602259.602266}. For the raster data, instead, we propose to use a modern compact data structure, the $k^2$-raster, which improves the performance of traditional methods.

The algorithms to manage independently each type of data and its corresponding data structure are well-known \cite{Manolopoulos:2005:RTA:1098699,k2rasterIS}. However, as explained, the algorithms to process both types of data jointly have been much less studied. Therefore, our proposal requires the design of new algorithms. In this work, we present  two new algorithms that are able to efficiently answer two operations having as input a vector dataset and a raster dataset. The first one is a spatial join between the two input datasets imposing a range restriction on the values of the raster dataset. The second algorithm obtains  the top-$K$ different objects of the vector dataset overlapping  the highest (or lowest) values of the raster dataset.

Our proposal obtains important savings in disk space, which are mainly due to the use of a $k^2$-raster for representing the raster data. In our experiments, the compressed raster data occupied between 9\% and 73\%  of the disk space needed by the original uncompressed raster data. 
However, the main contributions of this paper are the algorithms for  solving the aforementioned operations, which obtain savings also in main memory consumption and processing time.  Although the $k^2$-raster was designed to be used directly in compressed form, it is not trivial to save main memory while processing it. Thus, designing these algorithms becomes challenging, as the direct management of compressed data and indexes requires complex data structures and processes, which could negatively impact the results in main memory consumption and running time.\\ 

\section*{Related work}
\label{sec:relwork}

Spatial data can describe the space using two levels of abstraction. On the conceptual level, models describe the space using two different approaches: \textit{object-based spatial models} and \textit{field-based spatial models}~\cite{worboys2004gis}. It is in the logical level where spatial data models are divided into  \textit{vector models} and \textit{raster models}. 

Few data models consider the possibility of jointly using the object-based and the field-based spatial models. Even international standards separate both views~\cite{ISO19107,ISO19123}. The same situation can be found at the logical level, where international standards~\cite{OGCWFS,OGCWCS}  separate again both views and do not provide languages, data structures, or algorithms to perform queries that use information from both data models simultaneously.

Those geographical information systems that are capable of managing raster data are usually equipped with the operators of Map Algebra  \cite{tomlin1979mathematical,tomlin90}. Sometimes, as in the case of ArcGIS or GRASS, they support a variation of the zonal statistics operation  that, instead of receiving two input rasters, admit one vector dataset and one raster dataset. However, the vector dataset is first internally transformed into a raster dataset, such that a usual zonal statistics operation is executed over two raster datasets \cite{zonalS,zonalSG}.

The works in \cite{svensson1991geo,baumann1998multidimensional,vaisman2009multidimensional} provided the definition of  data models and query languages to  manage vector and raster data, but using a  different set of  query operators for each type of data.

 Grumbach et al.~\cite{grumbach2000manipulating} proposed a  data model  to represent vector and raster data with the same data abstraction. It includes spatial versions of  relational model operations like \emph{Projection}, \emph{Selection}, and \emph{Join}. These operations can manipulate vector and raster information without having to separate or distinguish the type of operands.

Brown et al.~\cite{brown2010overview} presented a data model that  represents vector and raster data with a  data abstraction based on multidimensional arrays. These works present data types, storage structures, and  operators to query vector and raster data, sometimes jointly, but unfortunately no details of implementation issues are provided (neither about the data structures nor the algorithms needed to support the model and the queries).

Corral et al.~\cite{corral1999algorithms} presented five algorithms for processing a join between a vector dataset, indexed with an R-tree, and a raster dataset, indexed with a linear region quadtree. In \cite{TPRtrees}, it is shown    an operation between regions and moving objects that obtains the predictive overlapping between them. A linear region quadtree is used again to index the raster dataset, whereas the predictive nature of the operation requires a different index for the vector data, namely a TPR*-tree. 
Unfortunately, these works tackled only  binary rasters, cells contain only black or white colours, and therefore they have a very limited real application.

In \cite{Eldawy:2017}, it was presented the {\em scanline} method, which is an algorithm to run the zonal statistics operation between a raster and a vector datasets without any previous transformation.

 Brisaboa et al. \cite{RodriguezBrisaboa17} presented a framework to store and manage  vector and compressed raster data, as well as an algorithm to solve a query that, given a vector and a raster dataset, returns the elements of  the vector dataset overlapping regions of the raster dataset that fulfill a range constraint. For example, having a vector dataset representing the neighbourhoods of a city and a raster storing the amount of nitrogen oxides in the air, a query could be ``return the neighbourhoods overlapping points where the concentration of nitrogen oxide is above 0.053 ppm''. However, their solution does not return the exact cells of the raster fulfilling the range constraint.
The vector dataset is indexed with an R-tree. The raster dataset is represented and indexed with a compact data structure called $k^2$-acc \cite{k2ones}, which needs a separate tree-based data structure  for each distinct value in the raster. More concretely, they use a compact data structure called $k^2$-tree \cite{ktree} for each  value. The $k^2$-tree is  a space- and time- efficient version of a region quadtree \cite{Klinger1971303,bb26519,Sam2006}, the typical index for binary raster data. To solve the query, the algorithm just requires the $k^2$-trees representing the values at the extremes of the range and the R-tree. The search starts at the root of the three trees and then proceeds in parallel a top-down traversal over all of them, pruning the branches of the trees when possible. 
The $k^2$-acc has two problems. First, it works well for range queries, that is, those specifying a range of values of the raster dataset (like the nitrogen oxide example just exposed); but obtains modest response times for other queries, such as obtaining the value of a given cell. The other problem of the $k^2$-acc concerns the size of the dataset. It is a compact data structure that gives good compression rates when the number of distinct values in the dataset is low. However, when the number of different values is large, the dataset occupies much more space than its uncompressed representation \cite{SSDBM16,k2rasterIS}, and scales really poorly when executing most queries.

  The framework proposed in this paper does not have any of these two problems of \cite{RodriguezBrisaboa17}, due to the use of a $k^2$-raster to represent raster datasets. Therefore, as the $k^2$-raster works well for all types of queries, and compresses the dataset even when the number of different values in the dataset is large, the framework achieves significant  savings in  query time, as well as, in space, in both disk  and main memory. In addition, the join operation included in our proposal differs from the one presented by Brisaboa et al., as our join algorithm also returns the cells of the raster dataset fulfilling the query constraints, that is, it is a \textit{real} join.

\section*{Background} \label{sec:k2raster}
In this section, we review the main techniques that will be used as basis of our proposal, and also some baselines that will be used to evaluate its performance.

Since it is a well-know data structure, we are not going to introduce here the R-tree \cite{Guttman:1984:RDI:602259.602266}, but we will include a brief explanation of the $k^2$-raster, the $k^2$-treap, and NetCDF to represent raster data.

\subsection*{$k^2$-raster}

The $k^2$-raster \cite{SSDBM16,k2rasterIS} is a compact data structure for storing an integer raster matrix in compressed form and, at the same time, indexing it. It consists in a compressed representation that allows fast queries over the raster data in little space. It efficiently supports queries such as retrieving the value of a specific cell or finding all cells containing values within a given range.

The $k^2$-raster exploits the uniformity of the integer matrix to obtain compression.
Following an analogous strategy to that of the $k^2$-tree \cite{ktree}, given a  raster matrix and a parameter $k$, the $k^2$-raster recursively subdivides the matrix into $k\times k$ submatrices and builds a conceptual tree representing these subdivisions and the minimum and maximum values of each submatrix. The subdivision stops in case that the minimum and maximum values contained in the submatrix are equal. This conceptual $k^2$-ary tree is then compactly represented using binary bitmaps and efficient encoding schemes for integer sequences. This leads to a compressed representation of the raster matrix with efficient indexing capabilities.

More concretely, let $n \times n$ be the size of the input matrix, being $n$ a power of $k$. To build the $k^2$-raster, it is necessary, first, to compute the minimum and maximum values of the matrix. If these values are different, they are stored in the root of a tree, and the matrix is subdivided into $k^2$ submatrices of size $n/k \times n/k$. Each of these submatrices generates a child node in the tree, where its minimum and maximum values are also stored. In case that these values are the same, the corresponding submatrix is not further subdivided. In case that these values are different, then this procedure continues recursively until the subdivision stops due to finding a uniform submatrix, where all the values are the same, or until no further subdivision is possible due to submatrices of size $1 \times 1$.

In case that the raster matrix is not squared or $n$ is not power of $k$, the matrix can be easily expanded to the squared matrix with size the following power of $k$, without imposing significant extra space.

\begin{figure}[!h]
	\begin{center}
		\includegraphics[width=1\textwidth]{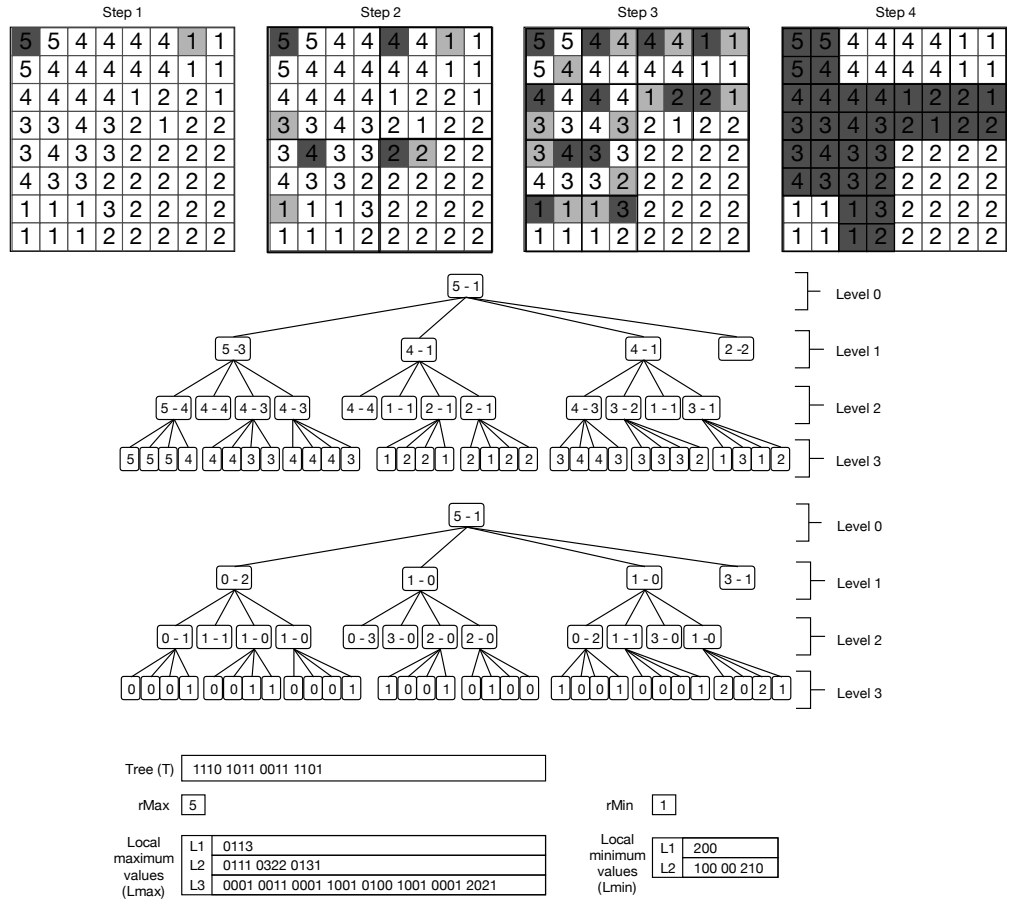}
	\end{center}
	\caption{{\bf $k^2$-raster example.} 
	Example of integer raster matrix (top), conceptual tree of the $k^2$-raster (centre-top), conceptual tree using differential encoding (centre-bottom), and final representation of the raster matrix using compact data structures (bottom). \textit{rMax} and \textit{rMin} denote the maximum and minimum values of the root node. \textit{Lmax} and \textit{Lmin} contain the maximum and minimum values of each node, following a level-wise order and using differential encoding. This example uses $k=2$.}
	\label{fig:k2raster}
\end{figure}

Fig \ref{fig:k2raster} shows an example of the recursive subdivision (top) and how the conceptual tree is built (centre-top), where the minimum and maximum values of each submatrix are stored at each node. The root node corresponds to the original raster matrix, nodes at level 1 of the tree correspond to submatrices of size $4\times 4$, and so on. The last level of the tree corresponds to cells of the original matrix. Notice, for instance, that all values of the bottom-right $4\times 4$ submatrix are equal; thus, its minimum and maximum values are equal, and it is not further subdivided. This is the reason why the last child of the root node has no children. 

To obtain a compressed representation of this conceptual tree, the $k^2$-raster uses two different approaches, one for representing the topology of the tree, and another for the maximum and minimum values stored at its nodes. On one hand, the shape of the tree is encoded using a bitmap, indicating in level-wise order whether a node has children or not. This corresponds to a simplified variant of LOUDS (\emph{level-ordered unary degree sequence}) tree representation \cite{Jac89}, which is a compact representation for trees. On the other hand, the maximum and minimum values are compactly encoded as the difference with respect to the maximum/minimum value stored at the parent node. These non-negative differences are stored as arrays, following the same level-wise order of the tree. The fact that the differences tend to be small is exploited using \emph{Directly Addressable Codes} (DACs) \cite{BLN13}, an encoding scheme for integer sequences that provides good compression and direct access to any given position.
At leaf nodes of the tree, only the maximum value is stored (as the minumum is the same).

We illustrate at Fig \ref{fig:k2raster} the final representation of the example matrix included at the top. In the centre-bottom part of the figure, we show the tree with the differences for the maximum and minimum values, whereas the data structures that compose the final representation of the $k^2$-raster are shown at the bottom part. Hence, the original  raster matrix is compactly stored using just a bitmap $T$, which represents the tree topology, and a series of compressed integer arrays, which contain the minimum and maximum values stored at the tree. Notice that when the raster matrix contains uniform areas, with large areas of equal or similar values, this information can be very compactly stored using differential and DACs encodings.

The $k^2$-raster not only obtains a very compressed representation of the raster matrix, but it also self-indexes the data, enabling fast queries over the raster matrix. These queries can be efficiently computed by navigating the $k^2$-raster; it is possible to simulate a top-down traversal over the conceptual tree by accessing the bitmap and the compact integer sequences in a very efficient way. In fact, some queries, such as finding cells having values within a specific range, can be answered faster with a $k^2$-raster representation than having the raster matrix in plain form, even when the $k^2$-raster requires much less memory space. 

It is possible to obtain better compression and navigation efficiency by using different $k$ values for each level. In particular, using just two values of $k$ works well in practice. This hybrid variant will be used in the experimental evaluation. It requires three parameters: $n_1$, $k_1$, and $k_2$, which indicate that $k=k_1$ for the first $n_1$ levels, and then $k=k_2$ for the rest.

As a summary, the $k^2$-raster joins three interesting characteristics in just one data structure: \textit{i)} it \textit{compactly stores the data}; \textit{ii)} the tree is \textit{a spatial index},  in fact it is built with the same procedure used by quadtrees, the typical index for raster datasets; and \textit{iii)} the minimum and maximum values stored at the nodes of the tree {\em index the  values stored at cells}. This last indexation is usually known as lightweight indexing, as it is inexpensive to offer \cite{SS17,Moe98,AA14}.

Compared to $k^2$-acc, the technique used in \cite{RodriguezBrisaboa17}, the $k^2$-raster not only obtains less space consumption and query performance, but it also scales better when increasing the size of the input data or when the raster matrix contains a large number of different values. 

\subsection*{$k^2$-treap}
The $k^2$-treap \cite{BdKNS16} is a data structure designed for answering fast top-$K$ queries over a grid of points, where the points have weights. It conceptually combines a $k^2$-tree with a treap data structure. Thus, thanks to the $k^2$-tree properties, it obtains compact spaces for representing the grid of points; and as it follows the ideas of a treap, it allows fast ranked queries.

The $k^2$-treap shares some common strategies with the $k^2$-raster. Conceptually, it is also a $k^2$-ary tree with some extra information in the nodes. Given a grid of points with weights (an integer matrix where some of the cells may contain no data), it locates the maximum value of the grid and stores this value along with its coordinates in the root node. Then, this maximum value is removed from its position, and the grid is subdivided into $k \times k$ submatrices. Each submatrix is represented in the tree as a child node, and the same procedure is repeated recursively. Leaf nodes in the $k^2$-treap represent submatrices where there are no more points with weights. 
This $k^2$-ary tree is also represented using compact data structures, including succinct bitmaps, differential encoding and DACs.
Top-$K$ queries over the grid are solved very fast, as the maximum values are indexed in the tree.

The $k^2$-raster and $k^2$-treap structures have not been compared before, as they have been designed for representing different types of data. The $k^2$-treap can also be used for representing raster data, where the grid is full of points with weights, and to solve access queries efficiently. However, the $k^2$-raster indexes the values of the raster better, as it stores not only the maximum, but also the minimum values, thus allowing us to search for cells with values in a specific range. Space requirements would depend on the uniformity of the raster matrix, as $k^2$-rasters can compact large areas of equal values, whereas $k^2$-treaps cannot exploit that property.

\subsection*{NetCDF} Network Common Data Form (NetCDF) \cite{lee2008netcdf} includes the data format and software libraries to compress, access and share array-oriented scientific data. More particularly, it can also be used for compressing raster matrices and allows accessing compressed datasets transparently without performing any explicit decompression procedure. Internally, 
NetCDF uses Deflate \cite{RFC1951}, which can be configured in ten compression levels. 
The compressed file is divided into blocks in such a way that when a portion of the raster is required, the library has to decompress one or more of those blocks.

NetCDF and $k^2$-raster have been compared recently \cite{k2rasterIS}. $k^2$-raster obtains compression ratios close to those achieved by NetCDF. NetCDF is faster than $k^2$-raster when accessing large portions of the data sequentially. On the other hand, $k^2$-raster obtains better access times to individual raster cells and when solving queries specifying conditions on the values of the raster. These queries are solved orders of magnitude faster, even compared with querying uncompressed NetCDF files, thanks to the indexing capabilities of the $k^2$-raster. 
Moreover, these two techniques follow different approaches, as NetCDF follows a classical disk-based approach, whereas $k^2$-raster is designed to operate completely in main memory.

\section*{A framework to store and process vector and raster data}\label{sec:framework}

In our framework,   the vector datasets are stored using a traditional setup indexed with R-trees,  and the raster datasets are stored and indexed using  $k^2$-rasters.
Next, we present two operations over that framework, which admit as input a vector dataset and a raster dataset.

\begin{figure}[!h]
	\begin{center}
		\includegraphics[width=1\textwidth]{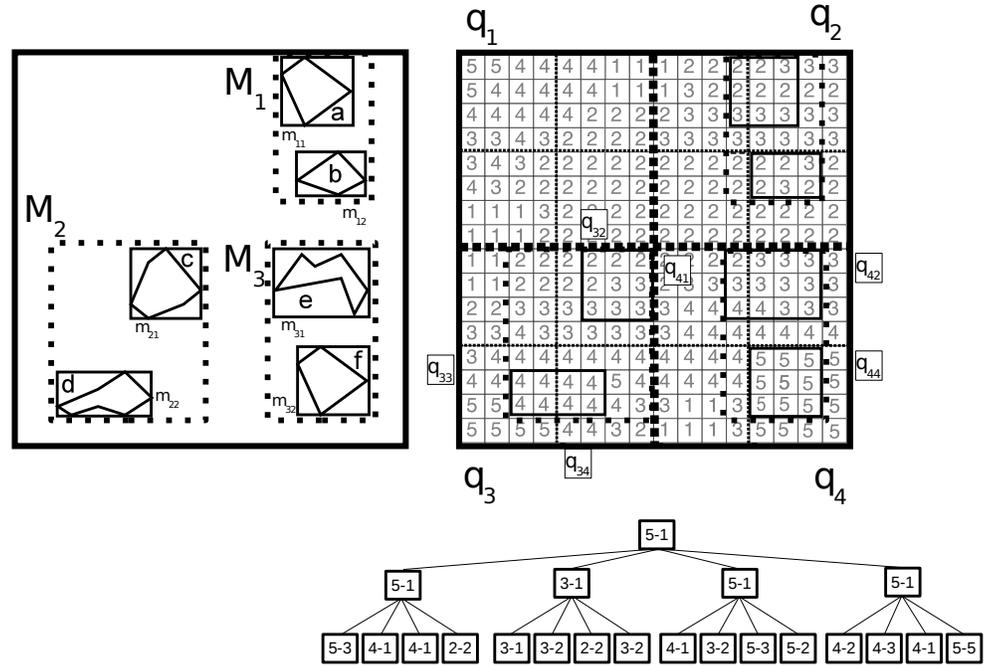}
	\end{center}
	\caption{{\bf A vector dataset (left) and a raster dataset (right).}
	A vector dataset (left), and its MBRs, and a raster dataset (right),  with the regions (quadrants) delimited by the divisions of the $k^2$-raster. For clarity,  the last level of the $k^2$-raster is omitted.}\label{running}
\end{figure}

Throughout the article, we will use the example of Fig \ref{running} to illustrate our explanations. The left part of the figure shows a vector dataset, and the right part shows a raster dataset. Using solid lines, the vector objects (labeled with lowercase letters) are depicted surrounded by their Minimum Bounded Rectangles (MBRs) that, for our example, are also the MBRs at the leaves of the R-tree indexing them. The MBRs  surrounded by rectangles with thick and very sparse lines ($M_1$, $M_2$, and $M_3$)  are the MBRs of the children of the root of the R-tree. 

Regarding the raster dataset at Fig \ref{running} (right), we show its cell values in light grey. The dotted lines are the regions of the space delimited by the splitting process of the $k^2$-raster (a hybrid one, $n_1=2$, $k_1=2$, and $k_2=4$). We call these regions {\it quadrants}. The thicker and densely dotted lines delimit the first level quadrants, denoted as  $q_1,$ $q_2,$ $q_3$, and $q_4$. The thinner and sparser dotted lines delimit the second level quadrants ($q_{11}, q_{12}, \ldots, q_{43}, q_{44}$). In the raster dataset, we also draw the MBRs of the vector objects,   with solid lines too, in order to easily see the overlays between the two datasets.
Under the raster, we also show the conceptual $k^2$-raster without the last level, since it just includes all the cells shown in the raster.

\subsection*{Basic definitions}

We will use the following notation during the next sections:

\begin{itemize}
    \item $p_r$ denotes a pointer to an R-tree node;
    \item $p_r.\mathit{MBR}$ denotes the MBR stored at node $p_r$;
    \item $p_r.\mathit{ref}$ denotes the list of pointers to the children of an internal node $p_r$, or the list of objects identifiers of a leaf node $p_r$;
    \item $p_k$ denotes a pointer to a  $k^2$-raster node;
    \item $p_k.\mathit{quad}$ denotes the quadrant corresponding to $p_k$;
    \item$p_k.\mathit{max}$ denotes the maximum value stored at $p_k$; and
    \item$p_k.\mathit{min}$ denotes the minimum value stored at $p_k$.
\end{itemize}

\subsection*{Efficient spatial join between raster and vector datasets} \label{sec:algorithm}

In this section, we present an algorithm to compute the join between a raster and a vector dataset, imposing a range constraint on the values of the raster.
Therefore, the algorithm returns the elements  of a vector dataset (polygons, lines, or points) and the position of the cells of the raster dataset
that overlap each other, such that the cells have values in a given range $[v_b,v_e]$. Formally, the query can be defined as:

\begin{Definition} \label{Def1}
Let $V$ be a set of vector objects, $R$ be the set of cells of a raster, and  $[v_b,v_e]$ a range of values. For any $O_u \in V$ and  $C_w \in R$, let  $O_u^{Sc}$  and $C_w^{Sc}$ be the spatial components of those elements and $C_w^{Val}$ the value stored at that cell. The join between the raster and the vector datasets, $V \Join R_{[v_b,v_e]}$, returns the set of tuples $\{ (O_1,\langle C_{1_1}^{Sc}, \ldots C_{1_{l1}}^{Sc}\rangle), (O_2,\langle C_{2_1}^{Sc}, \ldots C_{2_{l2}}^{Sc}\rangle), $ $\ldots,(O_n,\langle C_{n_1}^{Sc}, \ldots C_{n_{ln}}^{Sc}\rangle) \}$ such that $O_i^{Sc} \cap C_{i_y}^{Sc} \neq \emptyset$ and $v_b \leq C_{i_y}^{Val} \leq v_e$, for $1 \le i \le n, 1 \le y \le l_i$. 
\end{Definition}

That is, for each object of the vector dataset fulfilling the query constraints, the query returns the spatial component of the cells overlapping that object and having values in the queried range.

It is also possible to apply spatial restrictions on both datasets, that is, to restrict the join to windows or regions of the vector and the raster dataset.

In an index, handling directly  the exact geometries of indexed spatial objects would require complex and slow computations, thus making the index ineffective. Thus, 
when a query with spatial restrictions is run on a dataset indexed using MBRs, it is performed in two steps \cite[p.203]{Rigaux}. The first step, usually called {\it filter step}, retrieves the objects with MBRs fulfilling the constrains of the query. This step traverses the index applying the spatial constraints to the MBRs. The output can contain MBRs that satisfy the spatial constraints, whereas the exact geometry of the objects it contains does not. Then a second step, usually called {\it refinement step}, traverses the output of the filter step using the exact geometries of the objects to test if they  are actually part of the solution. This second step is costly but it is generally applied  over a much smaller number of objects. 
For the operation tackled in this section, we focus on the filter step.

In our algorithm, in order to reduce even more the burden of the refinement step, our filtering step separates all the selected MBRs into separate lists of {\it definitive results} and {\it probable results}.  Both lists include tuples formed by an object $O_i$ and the spatial component of the cells overlapped by the leaf MBR surrounding $O_i$ ($MBR_i$) that have values within the queried range.

\begin{itemize}
\item For any tuple of the definitive list, it holds that \textit{all} the cells overlapped by $MBR_i$ fulfill the range criterion, and thus, it is known for sure that $O_i$ is part of the final result (and does not need to be processed during the refinement step).

\item For any tuple of the probable list, it holds that only \textit{some} cells overlapped by $MBR_i$ fulfill the range criterion.  Therefore, the refinement step must be applied on $O_i$ to check if it is part of the final result. 

\end{itemize}

\subsubsection*{Checking the overlap}

The most critical operation of the join operation is to check whether a MBR of the vector dataset overlaps a region of the raster dataset having values within the queried range. This operation should be fast to obtain good running times. Therefore, our algorithm first tries a less accurate check (\textit{checkQuadrantJ}) that sometimes  is enough to make a decision. When it is not, the algorithm runs a second and more precise check (\textit{checkMBR}), which is obviously more expensive.

\textit{checkQuadrantJ}($p_r$, $p_k$, \textit{CellsRange}) receives as input a pointer $p_r$ to an R-tree node, a pointer $p_k$ to a $k^2$-raster node, and the range of values of the query.  It outputs a pair $( \mathit{typeOverlapQuad},  p_{k_{\mathit{deep}}})$, where $p_{k_{\mathit{deep}}}$ is a pointer to the deepest quadrant descendant of $p_{k}$, which completely contains $p_r.\mathit{MBR}$. The value of {\em typeOverlapQuad} depends exclusively on the raster values distribution inside the selected quadrant, and is one of the following:

\begin{itemize}

\item  {\em TotalOverlap} indicates that all the cells within $p_{k_{\mathit{deep}}}.\mathit{quad}$ have values in the queried range. Thus, in this case,  without any further check, the algorithm determines  that all the objects within $p_r.\mathit{MBR}$ are part of the solution and can be included in the definitive list.

\item {\em PossibleOverlap} indicates that $p_{k_{\mathit{deep}}}.\mathit{quad}$ contains some cells having values within the queried range, but also some cells with values outside that range. Therefore, it is not possible to make a decision at this point of the algorithm, and thus a  more thorough analysis is required.

\item  {\em NoOverlap} indicates that $p_{k_{\mathit{deep}}}.\mathit{quad}$ contains only cells of the raster that do not have values in the queried range. Thus, in this case, the algorithm can determine without any further check that all the objects within $p_r.\mathit{MBR}$ are not part of the solution, and thus, the subtree rooted at that node can be discarded.

\end{itemize}

\textit{checkMBR}($p_r$, $p_k$,\textit{CellsRange}) receives as input a pointer $p_r$ to a leaf R-tree node, a pointer $p_k$ to a $k^2$-raster node, and the range of values of the query. It returns only a variable {\em typeOverlapMBR} whose possible values are:

\begin{itemize}
	\item {\em TotalOverlap} indicates that all the quadrant cells overlapping $p_r.\mathit{MBR}$    have values in the queried range. Thus, this implies that all the objects within $p_r.\mathit{MBR}$ and the overlapping cells do not need to go through the refinement step, and therefore, they are included in the definitive list.

	\item  {\em PartialOverlap} indicates that the  geometry of  $p_r.\mathit{MBR}$ overlaps some quadrant cells having values in the queried range, but some others that do not. Thus, this implies that the objects within $p_r.\mathit{MBR}$ and the overlapping cells must go through the refinement step, and therefore, they are included in the probable list.
	
	\item {\em NoOverlap}  indicates that the exact geometry of  $p_r.\mathit{MBR}$ overlaps only cells of the raster quadrant that do not have values in the queried range. Therefore, all the objects within $p_r.\mathit{MBR}$ are not part of the result.

\end{itemize}

\textit{checkQuadrantJ} performs an preliminary less accurate  check  that, in some cases, allows the algorithm to make a decision. The key idea is that \textit{checkQuadrantJ} is very fast. It starts at the node of the $k^2$-raster provided as input, and then navigates the tree downwards selecting at each node the {\em unique} child that completely contains the MBR of the R-tree node, as long as the range of values delimited by the minimum and maximum values of the $k^2$-raster node intersect the query range. The navigation stops  when none of the children of the reached $k^2$-raster  node completely contains the MBR of the R-tree node, or when the minimum and maximum values of the $k^2$-raster node do not intersect the query range.

\textit{checkMBR} is the more accurate operation.
Obviously, this operation is more expensive, as it needs to navigate downwards {\em all} the children of the processed $k^2$-raster node that overlap the MBR of the R-tree node, until reaching  all the cells (leaves of the $k^2$-raster) overlapping the MBR.\\

\begin{example}
Let us take our running example at Fig \ref{running} to  illustrate these operations. Considering as input a pointer to the R-tree node whose MBR is $m_{12}$, a pointer to the root node of the $k^2$-raster, and the range of cell values [4,5], \textit{checkQuadrantJ} starts comparing the minimum and maximum values of the root ([1,5]) with the queried range ([4,5]). Since these ranges intersect, the navigation continues by checking if one of the children of the root of the $k^2$-raster completely contains $m_{12}$. This holds for $q_2$, and thus the process continues, again checking if the minimum and maximum values at that node ([1,3]) intersect the queried range ([4,5]). Since this is not the case, the operation ends outputting {\itshape typeOverlapQuad = NoOverlap}. This is a good example to see that it is possible to conclude that the content  of $m_{12}$ can be discarded already at the upper levels of the $k^2$-raster, in a fast way and without any further inspection of the input data. 

Now, let us consider \textit{checkQuadrantJ}  having as input the R-tree node corresponding to the leaf MBR $m_{22}$, the root node of the $k^2$-raster, and the queried range [4,5]. After checking the minimum and maximum values at the root of the $k^2$-raster,  the navigation goes to  $q_3$, the child that completely contains $m_{22}$. The maximum and minimum values at that  node ([1,5]) intersect the query range, but they are not fully within the queried range [4,5]. Since no smaller quadrant of $q_3$ completely contains $m_{22}$, the \textit{checkQuadrantJ} procedure ends outputting  {\itshape typeOverlapQuad = PossibleOverlap}. Therefore, the algorithm has to continue performing a deeper analysis using \textit{checkMBR} procedure, taking as input a pointer to the R-tree node of $m_{22}$ and a pointer to the $k^2$-raster node of $q_3$. Now the output is {\itshape typeOverlapMBR= TotalOverlap},  therefore all the objects inside $m_{22}$ and the overlapping cells are added to the definitive list.
This example shows that even when calling {\itshape checkMBR}, we take advantage from using the R-tree index, avoiding any further inspection of spatial objects inside some of the MBRs.

\end{example}

\subsubsection*{The algorithm}

Algorithm \ref{range} shows the pseudocode of the procedure that computes the filter step of the join of Definition \ref{Def1}. It receives as parameters a pointer to the root node of both the R-tree and the $k^2$-raster, and the query range of values of the cells of the raster. The two lists of definitive and probable results are declared in Line 1. Line 2 defines a stack used throughout the process, which keeps the  nodes of both trees to be processed. 

In Lines 3--4,  for each child of the root of the R-tree, the stack is initially filled with  a pair containing  a pointer to that child and a pointer to the root of the $k^2$-raster.  The {\it while} in Line 5 is the main loop of the algorithm, which in each iteration processes the top of the stack. As explained, it first tries a {\em checkQuadrantJ} operation. In  Line 8, a  {\it TotalOverlap} result allows the algorithm to make a decision and Lines 9--12 add the affected objects and cells to the definitive  list. 

In case {\em typeOverQuad} is \textit{PossibleOverlap}, the algorithm is forced to continue performing a deeper analysis. In Line 14, if the processed node of the R-tree is internal, the algorithm adds its children  to the stack, along with a pointer to the deepest quadrant containing the node ($p_{k_{\mathit{deep}}}$), and a new iteration of the main while starts. If the R-tree processed node is a leaf,  then a {\em checkMBR} call is issued. If the answer is {\it TotalOverlap}, the objects and the overlapping cells are added to the definitive list. A {\it PartialOverlap} implies the same addition but,  this time, to the probable list. In both cases, the {\em ExtractCells} method uses the $k^2$-raster to retrieve the coordinates of all the valid cells of $p_{k_{\mathit{deep}}}$ overlapping the MBR.

\begin{algorithm}
	\caption{\textbf{Join} ($p_{r\mathit{Root}}$, $p_{k\mathit{Root}}$, $[v_b,v_e]$)}
	\label{range}
		 Let \textit{Def} and \textit{Prob} be lists of tuples of the form $(O_i,\langle C_{i_1}^{Sc}, \ldots C_{i_{ln}}^{Sc}\rangle)$ \tcc*[h]{The lists of \textit{definitive} and \textit{probable} results}\\
		Let \textit{S} be a stack with pairs $(p_r,p_k)$ \tcc*[h]{$p_r$ is a pointer to an R-tree node and $p_k$ a pointer to a $k^2$-raster node}\\
		\ForAll{$p_{r\mathit{Child}}$ $\in$ $p_{r\mathit{Root}}.\mathit{ref}$ }{
		  push($S, ( p_{r\mathit{Child}},p_{k\mathit{Root}})$) \tcc*[h]{For each child of the root node of the R-tree, insert into the stack a pair with  pointers to that node and to the root of the $k^2$-raster }
		}
		\While{\textit{S} $\ne$ empty}{
		 $( p_r,p_k ) \leftarrow$ {\it pop}(S)\\
		 $( p_{k_{\mathit{deep}}}$,\textit{typeOverlapQuad}$) \leftarrow$ checkQuadrantJ($p_r, p_k,[v_b,v_e]$)\\
		\uIf {\textit{typeOverQuad} = \textit{TotalOverlap}}{
		\eIf{isLeafNode($p_r$)}{
		 addResult($p_r,$ ExtractCells$(p_r,p_{k_{\mathit{deep}}}), ~Def$) \tcc*[h]{Adds spatial objects and overlapping cells (having values in the queried range) to $\mathit{Def}$}}
		{
		  addDescendantsLeaves ($p_r.\mathit{ref}, p_{k_{\mathit{deep}}}, \mathit{Def}$) \tcc*[h]{Adds spatial objects and overlapping cells in descendant leaves to  $\mathit{Def}$}
		}
		}
		\ElseIf  {\textit{typeOverlapQuad} = \textit{PossibleOverlap}}{
		\eIf{isInternalNode($p_r$)}{
		\ForAll{$p_{r\mathit{Child}}$ $\in$ $p_{r}.\mathit{ref}$ }{
		 {\it push}(S, $( p_{r\mathit{Child}}, p_{k_{\mathit{deep}}})$)
		}
		}
		{
		 \textit{typeOverlapMBR}$\leftarrow$ checkMBR($p_r, p_{k_{\mathit{deep}}},[v_b,v_e]$)\\
		\uIf {\textit{typeOverlapMBR} = \textit{TotalOverlap}}{
		 addResult($p_r,$ ExtractCells$(p_r,p_{k_{\mathit{deep}}}), ~\mathit{Def}$) 
		}
		\ElseIf {\textit{typeOverlapMBR} = \textit{PartialOverlap}}{
		 addResult($p_r,$ ExtractCells$(p_r,p_{k_{\mathit{deep}}}), ~\mathit{Prob}$) 
		 }
		}
		}
		}
		\Return \textit{(Def,Prob)}
\end{algorithm}

As explained, the basic idea is to try to solve the query in the highest possible level of the two trees with the faster {\it CheckQuadrantJ}, and only when this is not possible, and we reach a leaf node, {\em checkMBR} is issued.

\begin{table}[t]
	\caption{Content of the stack during the example.}\label{stepsJ}
	\scriptsize
	\centering
	\begin{tabular}{|l|l|}
		\hline
		Step&\multicolumn{1}{c|}{Stack (S)}\\
		\hline
		1&$ ( M_1, q_{root}), ( M_2, q_{root}), ( M_3, q_{root})$\\
		2&$ ( M_2, q_{root}), ( M_3, q_{root})$\\
		3&$( m_{21}, q_3 ), ( m_{22}, q_3 ), ( M_3, q_{root})$\\
		4&$ ( M_3, q_{root})$\\
		5&$( m_{31}, q_4 ), ( m_{32}, q_4)$\\
		\hline
	\end{tabular}
\end{table}

\begin{example}
Using our running example and the query range [4,5], we are going to illustrate the  operation of the algorithm. The stack is initially filled with three pairs, each containing a pointer to the root of the $k^2$-raster and a pointer to one of the children of the root  of the R-tree, that is, to the nodes corresponding to $M_1$, $M_2$, and $M_3$ (see Step 1 of Table \ref{stepsJ}).

First, {\itshape checkQuadrantJ} is called (Line 7) with the top of the stack $( M_1, q_{root})$, which outputs $(\mathit{NoOverlap},q_2)$, given that the minimum-maximum values corresponding to $q_2$ ([1,3]) do not intersect [4,5]. Therefore, in this case, we can see one of the best cases for our algorithm, since it prunes a whole subtree rooted at one of the children of the root of the R-tree.

Then, the next top of the stack, $( M_2, q_{root})$, is processed. The {\itshape checkQuadrantJ} call returns $(\mathit{PossibleOverlap},q_3)$, and then, since $M_2$ is not a leaf,  Line 16 adds to the stack an entry,  for each of its children ($m_{21}, m_{22}$), with a pointer to that node and another to $q_3$ (see Step 3 of Table \ref{stepsJ}). 

Now, the next top of the stack, ($m_{21}$, $q_3$), is provided as input to {\itshape checkQuadrantJ}, which returns $(\mathit{NoOverlap}, q_{32})$, and then, it is discarded. With ($m_{22}$, $q_3$), {\itshape checkQuadrantJ} returns  
$(\mathit{PossibleOverlap}, q_3)$; observe that no child of $q_3$ completely contains $m_{22}$.  Therefore, since $m_{22}$ corresponds to a leaf, now the algorithm has to issue a {\em checkMBR} call, which returns a {\it TotalOverlap} value, and thus all the objects within $m_{22}$ and the overlapping cells are added to the definitive list.

The next top of the stack is $ ( M_3, q_{root})$, as shown in Step 4 of Table \ref{stepsJ}. {\em checkQuadrantJ} returns ($\mathit{PossibleOverlap}, q_4$), and then, since $M_3$ is not a leaf, Lines 14--16 push its children into the stack  producing the result shown in Step 5.

 {\itshape checkQuadrantJ}, with  ($m_{31}$, $q_4$) as input, outputs $(\mathit{PossibleOverlap}, q_4)$. So, a call to {\itshape checkMBR} is issued, which returns a {\it PartialOvelap}, and therefore, the objects inside $m_{31}$ and the overlapping cells having values in [4,5] are added to the probable list. The call to {\itshape checkQuadrantJ} with the last stack entry ($m_{32}$, $q_4$) returns $(\mathit{TotalOverlap}, q_{44})$, and thus the objects within $m_{32}$ are added to the definitive list. 
 \end{example}

\subsection*{Top-$K$ algorithm}

This query returns the \textit{K} objects  of a vector dataset that overlap  cells  of a raster dataset, such that the $K$ objects are those overlapping the highest (or lowest) cell values among all objects. Formally, we can define the top-$K$ query for the highest values as (the definition for the lowest values is analogous):

\label{sec:algorithmTop}

\begin{Definition} 
Let $V$ be a set of vector objects, and $R$ be the set of cells of a raster. For any $O_u \in V$ and  $C_w \in R$, let  $O_u^{Sc}$  and $C_w^{Sc}$ be the spatial components of those elements and $C_w^{Val}$ be the value stored at that cell.  The top-$K$ query $top_K(V,R)$ returns a set of $K$ tuples  $\{ (O_1, C_{l1}^{Val}), (O_2,C_{l2}^{Val}), $ $\ldots,(O_K,C_{lK}^{Val}) \}$,   such that $O_i^{Sc} \cap C_{li}^{Sc} \neq \emptyset$, $1\le i \le K$,  and $C_{li}^{Val} \geq C_j^{Val},$ for all pairs $( O_j,C_j^{Val})$ such that they are not part of in any tuple of $top_K(V,R)$ and  such that $O_j^{Sc} \cap C_j^{Sc} \neq \emptyset$ and, $O_j \in V$ and $C_j \in R$.  
\end{Definition}

An example of this query could be: let $Z$ be a region of the space, $R$ a raster dataset representing daily maximum temperatures in $Z$, and $V$ a vector dataset with polygons representing farms, distributed along $Z$. Then, a top-$K$ query could be ``\emph{Obtain the 10 farms in $V$ where the highest temperatures have been registered today}''.

\subsubsection*{Checking the overlap} 

As in the previous query, the  top-$K$ algorithm uses the same basic idea of \textit{checkQuadrantJ} and \textit{checkMBR}, but with some modifications. For instance, the two-step separation (filtering, refining) is no longer possible in this case, as we will see; and we need new versions for the check operations.

The new \textit{checkQuadrantT}  receives a pointer $p_r$ to an R-tree node and a pointer $p_k$ to a node of the $k^2$-raster and returns a pair $(  p_{k_{\mathit{deep}}}, max_{\mathit{deep}} )$. The component $p_{k_{\mathit{deep}}}$ is a pointer to the deepest descendant of  $p_k$ that completely contains $p_r.\mathit{MBR}$, and $max_{deep}$ is the max value stored at  $p_{k_{\mathit{deep}}}$. Observe that $max_{deep}$ is a tentative maximum value for the real maximum value in the  raster area overlapped by $p_r.\mathit{MBR}$. 
The more accurate check is now   \textit{checkGeometry}($p_r$, $p_k$), where $p_r$ is a leaf node. It  returns a list of tuples $( O_i,  C_{l_i}^{Val})$,   with one tuple for each object in $p_r.\mathit{ref}$, and where $C_{l_i}^{Val}$ is the value stored at the cells of the raster, among those overlapped by the object, that contain the maximum value.

Again, \textit{checkQuadrantT} is very fast, since it only checks the max/min values of the internal nodes of the $k^2$-raster. 
The operation \textit{checkGeometry} is more complex, because it obtains  
the raster portion  that intersects with $p_r.\mathit{MBR}$, and then, for each object in $p_r.\mathit{ref}$, a computational geometry algorithm is used to obtain the real maximum value overlapped by that object, and the cells having that value. Observe that the computational geometry algorithm is essential here, and can not be postponed to a later refinement step. We are not able to discard candidates based on the tentative max value identified by {\it checkQuadranT}, because it is possible that none of the objects inside $p_r.\mathit{MBR}$ overlaps with any of the cells with that value.

\subsubsection*{The algorithm}
Algorithm \ref{top} shows the pseudocode of the operation top-$K$, for the highest values; the version for the lowest values is obtained by simply changing the max values by min values and vice-versa. 

It receives, as input, pointers to the root of the R-tree and $k^2$-raster, along with the value of $K$. The algorithm processes both trees in parallel again. When traversing internal nodes, only when a uniform raster quadrant is processed, a decision can be taken, since it is sure that all objects within that quadrant overlap cells having the maximum/minimum value stored at the corresponding node of the $k^2$-raster.

If this is not enough to obtain $K$ objects, the algorithm finally reaches the leaf MBRs, and processes them in an order defined by the maximum value inside the raster region which they overlap. However, it is possible that only few, or even none, of the objects within each MBR overlap the cells with that maximum value.
Therefore, all those objects cannot still be added to the final result. They must be kept in a priority queue, and wait to be processed in the proper order, since the queue could still contain any pending object with a higher real maximum value, or any pending leaf MBR with a higher tentative value.

In the algorithm, the priority queue stores entries following the format $( vect, p_k, max, tent)$. {\em tent} is a flag to indicate whether the component {\em max} is tentative or a real value. Depending on the value of this flag,  the values of the rest of components are:
(i) {\em tent=true}, then \textit{vect} is a pointer to an R-tree node and $p_k$ is a pointer to a $k^2$-raster node; and
(ii) {\em tent=false}, then \textit{vect} is an object id and $p_k$ is a null value.

For each child of the root of the R-tree, Lines 3--5 add to the priority queue  a pointer to that child,  a pointer to the deepest node of the $k^2$-raster  that completely contains the MBR of that child, and the maximum value at that $k^2$-raster node, and thus, the {\em tent} flag is set to \textit{true}.

The  iteration of the {\em while} loop in Line 6 starts by checking the \textit{tent} component of  the head of the queue. If it is \textit{false} (Lines 9--10), then the algorithm extracts the head of the queue and  its values are added to the result.  Observe that a non tentative value means that the \textit{vect} component is the id of an object, whose exact geometry had been already checked before by the {\em checkGeometry} procedure.

If the {\em tent} value is {\em true}, then the algorithm checks if the pointer to the $k^2$-raster node points to a uniform area (the maximum and minimum values at the node are equal) in Line 13. In such a case,  without checking the exact geometry of the objects at the leaves of the subtree rooted at $p_{rq}$,  it is sure that those objects overlap cells with the retrieved $max_q$ value, and thus they are added to the result (up to $K$ entries) with the  procedure {\em addDescendantsLeaves}, which traverses the R-tree downwards to retrieve them.

If the quadrant is not uniform, the algorithm checks  if the pointer to the R-tree is a leaf or not (Line 15). If it is a leaf, since the max value obtained from the queue was tentative, the algorithm performs a call to {\em checkGeometry}. That call returns a list of tuples $( O_i,  C_{l_i}^{Val})$, with one tuple for each polygon in the R-tree leaf. For each of those tuples, if the real maximum is equal to the tentative maximum value (Line 18), then that object is added to the response. If it is lower (Line 23), then it is added back to the priority queue, but now with the format $( O_i, null, C_{l_i}^{Val}, false)$. If the pointer to the R-tree is not a leaf (Line 25), the algorithm adds to the priority queue each of the children, along with the deepest quadrant that completely contains it, along with its tentative maximum value.\\

\begin{algorithm}
 
	\caption{\textbf{Top-\boldmath$K$} ($p_{r\mathit{Root}}$, $p_{k\mathit{Root}}, K$)}
	\label{top}
		 Let \textit{T and L}  be lists of elements  $( O_u, C_{l_u}^{Val})$ \tcc*[h]{T holds the output of the algorithm and L is a auxiliary list.}\\
		 Let \textit{Q} be a priority queue with entries $( vect,p_k,max, tent )$  \\
		\ForAll{$p_{r\mathit{Ref}}$ $\in$ $p_{r\mathit{Root}}.\mathit{ref}$ }{
		 $( max_{deep}, p_{k_{\mathit{deep}}}) \leftarrow$ checkQuadrantT($p_{r\mathit{Ref}}$,$p_{k\mathit{Root}}$)
		  insert($Q, ( p_{r\mathit{Ref}},p_{k_{\mathit{deep}}}, max_{deep}, true )$)\tcc*[h]{Inserts in the priority queue   each child of the root node of the R-tree}
		}
		\While{\textit{Q} $\ne$ \textit{empty} and \textit{sizeOf(T)}$<$\textit{K}}{
		 $tent \leftarrow tent(head(Q))$ \tcc*[h]{Obtains the \textit{tent} flag of the head of the  queue}\\
		\eIf {tent=false}{
		  $( O_q, null, max_q ,tent) \leftarrow$ {\it head}(Q) \tcc*[h]{Extracts  the head of the queue}      \\        add(T, $( O_q, max_q)$)  \tcc*[h]{The max value at the priority queue is real, add to the result} }
		{ 
		 $( p_{rq},p_{kq}, max_q ,tent) \leftarrow$ {\it head}(Q)\\
		\uIf{$p_{kq}.min=p_{kq}.max$}{
		 $addDescendantsLeaves(T, p_{rq},p_{kq}, max_q)$\tcc*[h]{A uniform quadrant, all descendants can be added to the result}}
		\uElseIf{isLeafNode($p_{rq}$)}{
		 $L \leftarrow$ checkGeometry($p_{rq}, p_{kq}$)  \tcc*[h]{Obtains a list of tuples} \\
		\ForAll{ $( O_i,  C_{l_i}^{Val}) \in  L$ }{
		\eIf {$C_{l_i}^{Val}=max_q$}{
		 add(T, $( O_i, max_q)$) \tcc*[h]{The real value is equal to the tentative, then add to the output}\\
		\lIf{\textit{sizeOf(T)=K}} {
		 break}
		}
		{ insert($Q, ( O_i, null, C_{l_i}^{Val}, false )$) \tcc*[h]{The real max is smaller than the tentative one; the object is inserted in the queue}
		} }
		}
		\Else{\ForAll{$p_{r\mathit{Ref}}$ $\in$ $p_{rq}.\mathit{ref}$ }{
		$( max_{deep}, p_{k_{\mathit{deep}}}) \leftarrow$ checkQuadrantT($p_{r\mathit{Ref}},p_{kq}$)\tcc*[h]{It descends one level of the R-tree, and adapts the nodes of the $k^2$-raster accordingly}
		 insert(Q, $( p_{r\mathit{Ref}}, p_{k_{\mathit{deep}}},  max_{\mathit{deep}}, true)$)
		}
		}
		}
		}
		\Return $T$
\end{algorithm}

\begin{table}[t]
	\caption{Content of the priority queue during the computation of a \textit{top-1} query over the example.}\label{steps}
	\scriptsize
	\centering
	\begin{tabular}{|l|l|}
		\hline
		Step&\multicolumn{1}{c|}{Priority queue (Q)}\\
		\hline
		1&$ ( M_2, q_3, 5, true), ( M_3, q_4, 5, true), ( M_1, q_2, 3, true)$\\
		2&$ ( m_{22}, q_{3}, 5, true), ( M_3, q_4, 5, true), ( m_{21}, q_{32}, 3, true), ( M_1, q_2, 3, true)$\\
		3&$( M_3, q_4, 5, true), ( d, null, 4, false ), ( m_{21}, q_{32}, 3, true), ( M_1, q_2, 3, true)$\\
		4&$( m_{31}, q_{4}, 5, true), ( m_{32}, q_{44}, 5, true), ( d, null, 4, false ), ( m_{21}, q_{32}, 3, true), ( M_1, q_2, 3, true)$\\
		5&$ ( m_{32}, q_{44}, 5, true),  ( d, null, 4, false ), ( e, null , 3, false ),( m_{21}, q_{32}, 3, true), ( M_1, q_2, 3, true)$\\
		\hline
	\end{tabular}
\end{table}

\begin{example}
Let us illustrate the algorithm computing a \textit{top-1} query over the example of Fig \ref{running}. 
Lines 3--5 add each child of the root of the R-tree to the priority queue. The first row of Table \ref{steps} shows the queue content after this first step. 

Line 7 obtains the {\itshape tent} value of the head, and since it is {\em true},  Line 12 extracts the head $( M_2, q_3, 5, true)$; since $q_3$ is not uniform (the maximum and minimum values of $q_3$ are different), and $M_2$ is not a leaf, the flow reaches Line 25 and then,  the children of $M_2$ are added to the queue (see Step 2). Observe that, for each child, a  call to {\itshape checkQuadrantT} provides the deepest quadrant that completely contains that child and the maximum value at that quadrant.

In the next iteration of the \textit{while} loop,  given that the head $( m_{22}, q_{3}, 5, true)$ has a tentative maximum, $q_3$ is not uniform, and $m_{22}$ is a leaf, then the flow reaches Line 16 and $checkGeometry(m_{22}, q_{3})$ is issued. The response is   $( d,  4 )$, and thus, the \textit{for} loop of Line 17 only checks that entry; and since the real maximum is smaller than the tentative one (Line 18), the tuple $( d, null, 4, false )$ is added to the queue in Line 23 (see Step 3). Observe that,  with a tentative maximum value of 5, which is greater than 4, $M_3$ is still waiting in the queue, since it could contain objects overlapping cells with value 5, that would be added to the final result before $d$.

In fact, the next iteration processes $( M_3, q_4, 5, true)$. Since {\itshape tent} is true, $q_4$ is not uniform, and $M_3$ is not a leaf, the flow reaches Line 26, and then the children of $M_3$ are added to the queue (Step 4).

The next dequeued head $( m_{31}, q_{4}, 5, true)$ produces  a call to \textit{checkGeometry} that  returns a real maximum value (3) smaller than the tentative one, and thus $( e, null , 3, false )$ is added to the queue (see Step 5).

Finally, the next iteration processes $( m_{32}, q_{44}, 5, true)$. After dequeuing it, since {\itshape tent} is true, Line 14 checks the maximum and minimum values at  $q_{44}$; since they are equal (it is a uniform quadrant),  a call to {\itshape addDescendantsLeaves} adds the object in $m_{32}$ to the output (in this case the polygon {\itshape f}). This ends the computation of top-1.
\end{example}

\section*{A note on complexity}
\subsection*{Time}

Observe that our two algorithms can be modeled as a spatial join between an R-tree and a raster where the spatial predicate is the intersection of areas or zones.
 When using spatial data structures as, for example, the R-tree and quadtrees, the worst-case analytical complexity for window queries, which include among others spatial joins, does not reflect what really occurs in practice, because the behaviour depends on the size of the window and on the intersection area or zones at each level of the spatial structure \cite{B1993}. Recall that the $k^2$-raster is conceptually a quadtree.

Therefore, to better predict the actual performance, the usual approach is to develop cost models. For example, Theodoridis et al. \cite{T2000} presented a cost model to approach the cost of a window query using an R-tree. 
On the other hand, Theodoridis et al. \cite{Theodoridis:1998:CMJ:645483.656214} and Corral et al. \cite{Corral2006} proposed cost models that estimate the time required to calculate a spatial join between two R-trees considering a spatial predicate and distance, respectively.

The development of a cost model for our algorithms is beyond the scope of this paper, becoming a result by itself. However, we present  a preliminary analysis.

Let $\ell$ be the number of MBRs in the last level of the R-tree and let $n$ be the number of rows/columns of the raster. 
A simple   
baseline approach to solve the spatial join would proceed first by obtaining the MBRs in the leaves of the R-tree, and then, 
for each one, inspect all cells of the raster that overlap that MBR. The number of cells that overlap an
MBR is bounded by the size of the raster, that is, $O(n^2)$. Therefore the time complexity of that 
baseline approach is $\Theta(\ell  n^2)$.

Now, we analyze the spatial join in our framework.  
The advantage of our algorithms is that they do not need to always reach the last level of the R-tree to obtain the output. There are  several cases where the search from the root of the R-tree stops before the last level:

\begin{itemize}
    
\item In the case of Algorithm \ref{range}:
\begin{enumerate}
    \item When all the values of the region contain the exact same value, that is, when the minimum and maximum values of the $k^2$-raster node are equal, and that value is not within the queried range.
    \item When the maximum value of the processed $k^2$-raster node is smaller than the left extreme of the queried range, or when the minimum is greater than the right extreme of the queried range.
\end{enumerate}

\item In the case of Algorithm \ref{top}, in addition to the previous cases, the algorithm stops when the size of the resulting set reaches the desired $K$.
\end{itemize}

The worst case for Algorithm \ref{range} corresponds to that where for all the MBRs of the $R$-tree, we need to traverse all the nodes of the $k^2$-raster. Considering the $\ell$ MBRs and the number of nodes of a complete $k^2$-raster $\sum_{i=0}^{h}{k^{2i}}=\frac{k^{2h+1}-1}{k^2-1} \approx k^{2h}=O(n^2)$, where $h=O(log_k n)$ is the height of the $k^2$-raster, then the cost for the worst case is $O(\ell n^2)$.

In the case of Algorithm \ref{top}, the predominant costs are those corresponding to the management of the priority queue $Q$ and the cost of function {\em checkGeometry}. Let us simplify the analysis considering just top-1 query ($K=1$). On one hand, the worst case would require that all MBRs need to be inserted in $Q$, as the maximum value is guiding the search to areas close to (but not within) all the other MBRs different for the top-1. This causes that the total cost of inserting and obtaining the head of the queue at each step is $O(\ell \log_2 \ell)$. On the other hand, {\em checkGeometry} requires a verification, in the worst case, of $O(n^2)$ cells. Thus, the total cost for Algorithm  \ref{top} is $O(\ell \log_2 \ell) + O( \ell n^2) = O( \ell n^2)$. It is noticeable that, for the worst case, the overall cost of top-$K$ does not depend on $K \le \ell$, and part of the $O(\ell)$ inserted elements in $Q$ will be members of the solution and will be also inserted into list $T$.

Thus, the overall cost for the worst-case scenario of our algorithms is the same as that of the baseline. However, it is a very pessimistic analysis that is not considering the amount of searches that are not reaching the last level of the R-tree. 
Let us consider the MBRs in the last level of the R-tree such that in some step of the tree traversal, the search ended before reaching them, and denote $R$ the percentage of those MBRs. 
Therefore, the question is the size of $R$; that study would require a cost model, however, as both our experimental results and previous work using lightweight indexation (which include min, max values at nodes) \cite{SS17,Moe98,AA14} show, the real performance is improved.

\subsection*{Storage}

The space consumption for the case of the R-tree was presented in the original work \cite{Guttman:1984:RDI:602259.602266}. In the case of the $k^2$-raster, it has not been presented previously, again, because a worst case complexity does not fairly reflect the behaviour of the data structure.

However, we can compare the worst case of a baseline approach and that of the $k^2$-raster. Being  ${\cal M}$  the highest value of the raster, a simple baseline that stores each cell using $ \lceil \log  {\cal M} \rceil$ bits   requires in total $ N_b= \lceil \log {\cal M} \rceil n^2 = \Theta(\lceil \log {\cal M} \rceil n^2)$ bits.

The most unfavorable scenario for the $k^2$-raster occurs when all cell values are different. In this case,
 it has to store the topology of the tree (bitmap $T$) and the two sequences of minimum and maximum values for each level of the tree, except for the last level, where only the sequence of maximum values is stored. Assuming $h$ the height of the $k^2$-raster, the number of nodes at each level $i$, in a complete tree is $k^{2i}$, being $0 \le i \le h$. 
 The number of internal nodes of a complete $k^2$-raster is $\sum_{i=0}^{h-1}{k^{2i}}=\frac{k^{2h}-1}{k^2-1} \approx k^{2h-1}=O(n^2)$ and the number of leaf nodes $k^{2h}$.
 
 Therefore the required storage is $N_{r}= \frac{k^{2h}-1}{k^2-1} + \sum_{i=0}^{h-1}(c(min_1, min_2, \ldots, min_{k^{2i}})+ c(max_1, max_2, \ldots, max_{k^{2i}}))+ c(max_1, max_2, \ldots, max_{k^{2h}})$, where $c(.)$ is the number of bits required by DACs encoding.
 Since the internal nodes have two lists and the last level only one, the size of these lists joined is $s \approx 2k^{2h}$. According to \cite{BLN13}, the number of bits of a compressed sequence of length $s$ with DACs is $N_0 + 2s\sqrt{\frac{N_0}{s}}$, where $N_0=\sum_{j=1}^{s}{\lfloor \log x_j \rfloor +1} \le  \lceil \log {\cal M} \rceil 2k^{2h} = O(\lceil \log {\cal M}\rceil n^2)$ bits, thus the worst case space of the $k^2$-raster is $N_r = O(\lceil \log {\cal M} \rceil n^2)$.

Again the worst case is the same as that of the baseline. 
However, the baseline does not benefit from Tobler's law, thus the total cost will be exactly $\lceil \log {\cal M} \rceil n^2$ bits always. Nevertheless, in practice, the $k^2$-raster  can reduce significantly the storage due to two effects:

\begin{enumerate}
    \item Parts of the raster having the same value are simply represented by only one number. This is a feasible assumption due to Tobler's law.
    \item The numbers are represented using difference encoding with respect to the maximum value of the parent node, which decreases the magnitude of the numbers when increasing the depth in the tree. In other words, there is a new maximum value ${\cal M}_i, ~0\leq i \leq h$ for each level of the tree and, it is sure that ${\cal M}_{h} < {\cal M}$ and ${\cal M}_{i} \leq {\cal M}_{i-1} ~1 \leq i \leq h$. 
    
    This reduction of the magnitude of the numbers is exploited by DACs encoding, thus obtaining further compression.
\end{enumerate}

Again, to determine the percentage of raster cells that are compacted in just one number (uniform areas) and the effective reduction of magnitudes for the maximum-minimum sequences, better analysis can be obtained if a model of the data is considered.

\section*{Experimental evaluation}\label{sec:experiments}

\subsection*{Experimental Framework}\label{sec:ExperimentalFramework}
In this section, we present a set of experiments to measure the space consumption and processing time of the two algorithms.

The machine used in the experiments was equipped with an Intel\textsuperscript{\textregistered} Core\textsuperscript{TM} i7-3820 CPU @ 3.60 GHz (4 cores) processor and 64 GB of RAM. The operating system was   Ubuntu 12.04.5 LTS with kernel 3.2.0-115 (64 bits). All the programs were coded in C++ and  compiled using  gcc version 4.6.4 with \texttt{-O3}
options. 

Time is measured using the sum of {\sc user} and {\sc sys} values   provided by the Linux {\sc time} command, in seconds, except in the experiments of cold start that show the {\sc real} value. The memory consumption is measured taking the {\sc vmpeak} value of the pseudo-file system  {\sc proc},  in megabytes. 

\tolerance 1000000 \pretolerance 1000000
We obtained the R-tree implementation from \url{https://libspatialindex.github.io} and we configured it with a page size of 4 KB and a fill factor  of 70\%. We used  the authors' implementation of the $k^2$-raster,  available at \texttt{http://lbd.udc.es/research/k2-raster/}. We chose the heuristic variant \cite{k2rasterIS} with a hybrid configuration using $n_1=4$, $k_1=4$, and $k_2=2$. The bitmaps needed by the $k^2$-raster code were arranged with an implementation of the {\it rank} operation that requires a  5\% of extra space on top of the bitmap \cite{GGMN05}. In addition, the code uses also a DACs implementation configured to obtain the maximum compression  restricting the maximum number of levels to 3. 
\tolerance 500 \pretolerance 500

\subsection*{Baselines}

There is no previous implementation for the spatial join operation in the literature. The closest related works are those by Corral et al.~\cite{corral1999algorithms} and Brisaboa et al.~\cite{RodriguezBrisaboa17}. 
In the case of the top-$K$ operation, there is not even any close related work. 

The work of Corral et al.\ only considers binary rasters, thus we run a separate experiment for this software, as our algorithm is designed for rasters with different values, which implies more complex data structures.

The work of Brisaboa et al.\ returns only the vector objects, but not the raster cells. However, we modified the authors' algorithm in order to obtain the same output as ours. We made our implementation using the same R-tree used for the rest.  The data structure for storing the raster data, called $k^2$-acc, was the authors' implementation.

Furthermore, we  programmed three additional baselines. Two of them load the complete  uncompressed dataset into main memory, where the data are kept as a simple array following  a simple filling  curve row by row. One is denoted \texttt{Plain-Ints} and uses a 32-bit integer representation for each cell value. The other is termed  \texttt{Plain-Bits} and uses $\lceil log(\#v) \rceil$ bits for each cell value, being $\#v$ the number of different values in the original matrix.  Finally, the third baseline uses the NetCDF libraries from Unidata\textsuperscript{\textregistered}, available at \url{http://www.unidata.ucar.edu/downloads/netcdf/}, to compress and decompress raster data, with the recommended \emph{deflate} level 2.

For the baselines, we use  two naive strategies: 

\begin{enumerate}[I.]
    \item  The strategy labeled \texttt{mbrs} looks for all leaf MBRs in the R-tree and then overlaps them with the raster dataset, where, for each intersection between an MBR and the overlapping portion of the raster: 
    \begin{itemize}
        \item In the case of the join, it searches the cells having values in the queried range.
        \item  In the case of the top-$K$ query, it obtains the largest cell value, and it retains those MBRs with the highest values.
    \end{itemize}
    
   \item The strategy labeled \texttt{cells} starts by the other side, that is:
   \begin{itemize}
   \item In the case of the join, it obtains the cells of the raster that meet the range specified in the query, then those positions are checked using the R-tree to see if they lie within an MBR.
    
   \item The cells with the top-$K$ values (not necessarily different) in the raster dataset are obtained. Then the R-tree is used to look for overlapping MBRs. If not enough MBRs are found, the next top-$K$ values in the raster are obtained, and so on. 
   \end{itemize}
\end{enumerate}

In the case of the join, for \texttt{Plain-Ints}, \texttt{Plain-Bits}, and \texttt{NetCDF}, we only included the \texttt{mbrs} method, as the \texttt{cells} approach was too slow. However, we include the \texttt{cells} version using a $k^2$-raster representation (\texttt{k$^2$-raster-cells}), in order to see the difference with our algorithm. 

For the  top-$K$ query, we replaced \texttt{Plain-Bits} by \texttt{k$^2$-treap}, which is a baseline based on $k^2$-treap that uses the implementation published by the Database Lab research group, available at \url{http://lbd.udc.es/research/aggregatedRQ/}. The reason is that this structure, although it is very inefficient for other types of queries, including the join operation,  was specifically designed for the top-$K$ query.   There is just one implementation for \texttt{k$^2$-treap}, since strategy \texttt{cells} seems the best option in that scenario (as it has an index that allows us to rapidly obtain the top-$K$ cells). The code of all these baselines is available at \url{https://gitlab.lbd.org.es/fsilva/basic_raster}.

 Another possible comparison  is the traditional approach of translating the vector dataset to raster, and then run a classical raster-raster algorithm. However, that procedure is costly, for example, we used Grass (\url{https://grass.osgeo.org/}) latest version (7.4) to translate our two vector datasets (see next section). This took  6 and 1.8 seconds, which is much more than the time needed by our algorithm to run the query, and  therefore we decided not to include this baseline in our experiments.

\subsection*{Datasets}
The datasets used in our experiments, both the vector and the raster datasets, correspond to real data that were scaled to fit in the same space. Notice that this is the most demanding scenario for a spatial join, as it is not possible to exclude any branch of the R-tree. We describe them more in depth in the following section.

\subsubsection*{Raster datasets}

\begin{table}
	\footnotesize
	\centering
	\caption{Raster dataset for Scenario I. Values in Megabytes.}\label{table:datasetsCAT}
	\begin{tabular}{|l|r|r|r|r|r|r|r|r|}
		\hline
		&    &      & \# dif. & \texttt{Plain-}& \texttt{Plain-}& &&\\
		Name & \#rows & \#cols & values & \multicolumn{1}{c|}{\texttt{Ints}}  & \multicolumn{1}{c|}{\texttt{Bits}} &\texttt{k$^2$-raster} &\texttt{NetCDF}&\texttt{k$^2$-treap}\\
		\hline
		\texttt{DTM-1$\times$1}	&	4,100	&	5,849	&	868	& 91 & 30 & 10	& 10&15\\	
		\hline
		\texttt{DTM-2$\times$2}	&	8,242	&	11,737	&	1,201 & 369	& 126 & 38&38&	58\\
		\hline
		\texttt{DTM-3$\times$3}	&	12,403	&	17,643	&	1,503	&	835 & 296 & 81&84&132 	\\
		\hline
		\texttt{DTM-4$\times$4}		&	16,564	&	23,564	&	1,761	& 1,479 & 535 &	143&151&236\\
		\hline
	\end{tabular}
\end{table}

\begin{table}
	\footnotesize
	\centering
	\caption{Raster dataset for Scenario II. Values in Megabytes.}\label{table:datasets0533}
	\begin{tabular}{|l|r|r|r|r|r|r|r|r|}
		\hline
		&    &      & \# dif. & \multicolumn{1}{l|}{\texttt{Plain-}}& \multicolumn{1}{l|}{\texttt{Plain-}}& &&\\
		Name & \#rows & \#cols & values &\multicolumn{1}{c|}{\texttt{Ints}}  &\multicolumn{1}{c|}{\texttt{Bits}}   &\texttt{k$^2$-raster} &\texttt{NetCDF}&\texttt{k$^2$-treap}\\		\hline
		\texttt{DTM-4$\times$4$_{0}$}& 16,564	&	23,564 & 2,153		&	1479 & 555 &	139 &147 & 230	\\
		\hline
		\texttt{DTM-4$\times$4$_{1}$} & 16,564	&	23,564 & 21,491	&	1479	& 693 & 362&322	& 361\\
		\hline
		\texttt{DTM-4$\times$4$_{2}$}& 16,564	&	23,564 & 208,493&	1479& 832 &	540 &454& 509	\\
		\hline
		\texttt{DTM-4$\times$4$_{3}$}& 16,564	&	23,564 & 1,829,334&	1479	& 970 &	708&611& 659	\\
		\hline
	\end{tabular}
\end{table}

The raster datasets were obtained from Spanish Geographic Institute (SGI), which includes several DTM (Digital Terrain Model) data files with the spatial elevation data of the terrain of Spain. The complete DTM of Spain is divided in several tiles, and each tile is stored in a separate file.
Each file contains a grid of regularly spaced points, with 5 metres of spatial resolution, storing their spatial elevation as real numbers of at most 3 decimal digits. 

In order to evaluate the performance of our algorithms, we obtained two different sets of rasters, which serve us to analyze the scalability of our algorithms in two different scenarios:

\begin{itemize}
	\item Scenario I: designed for analysis based on raster size. Table \ref{table:datasetsCAT} shows the details of our set  of rasters   of increasing size.
	The set is  formed by four  collections, each one containing different  matrices of the same size. 
	Our initial collection, \texttt{DTM-1$\times$1}, was built selecting 25 samples of exactly one tile of the DTM; the collection \texttt{DTM-2$\times$2} was built using 2$\times$2 adjacent tiles, and so on. The data shown in the table (e.g. the number of different values) correspond to the mean values obtained for all the matrices in each collection. We will report the average values for space and time results, in the experiments, for each collection, thus, avoiding the dependence on the selection of a unique matrix. In this scenario, we only considered the integer part of the value stored at each cell of the matrix.
	
	\item Scenario II: 
		designed for analysis based on the distribution of raster values. Table \ref{table:datasets0533} shows the details of the set of rasters that form a scenario of equal sized datasets, but with increasing number of distinct values.
	We chose one of the \texttt{DTM-4$\times$4} datasets, and generated a collection of matrices varying the number of different values. 
	For this, we truncated the original values by taking 0, 1, 2, and 3 decimal digits.

\end{itemize}

All measures are the mean resulting from running 10 queries with random query ranges over each dataset for each collection. That means that, for example, in the \texttt{DTM-1$\times$1} collection, $10 \times 25=250$  queries were run. All queries are available at: \url{http://lbd.udc.es/research/k2-raster/}

Tables \ref{table:datasetsCAT} and \ref{table:datasets0533} show that by using \texttt{k$^2$-raster}, our framework obtains important savings in disk space with respect to the uncompressed representations. In the datasets of Scenario I, \texttt{k$^2$-raster} occupies around 10\% of the space occupied by \texttt{Plain-Ints} and between 27\% and 33\% of that occupied by \texttt{Plain-Bits}. For Scenario II, \texttt{k$^2$-raster} occupies 
between 9\% and 48\% of the space occupied by \texttt{Plain-Ints} and between 25\% and 73\% of that occupied by \texttt{Plain-Bits}. 

With respect to the compressed representations, \texttt{k$^2$-raster} obtains similar results to those of \texttt{NetCDF}, except in the datasets with many different values of Scenario II, where \texttt{NetCDF} occupies around 85\% of the space required by \texttt{k$^2$-raster}. Finally,  \texttt{k$^2$-raster} occupies around 65\% of the space used by \texttt{k$^2$-treap}, except in the datasets with many distinct values, where they  are approximately on a par.

\subsubsection*{Vector datasets}

\begin{figure}[!h]
	\centering
		\includegraphics[width=1\textwidth]{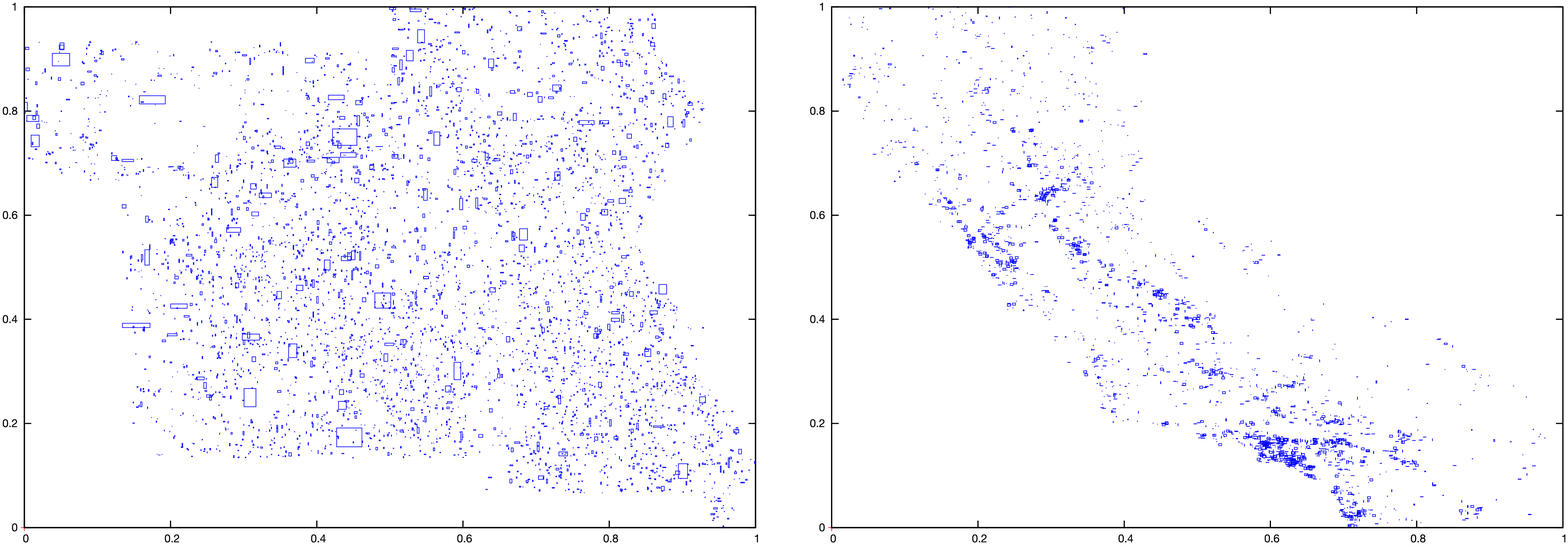}
	\caption{{\bf Spatial  distribution of the MBRs of the vector datasets \texttt{vects} (left) and \texttt{vecca} (right).}}
	\label{fig:rasterdatasets}
\end{figure}

We obtained two datasets from the ChoroChronos.org web site (\url{http://www.chorochronos.org/}). In our experiments, the label \texttt{vects}  refers to the dataset \emph{Tiger Streams} and the label  \texttt{vecca} refers to the dataset  \emph{California Roads} of that site. These two datasets have a very different number of MBRs and spatial distribution, as shown in Fig \ref{fig:rasterdatasets}. More concretley, \texttt{vects} contains 194,971 MBRs and \texttt{vecca} contains 2,249,727 MBRs.

\subsection*{Spatial Join}

Before presenting the results of our experiment, we must make a point. The three baselines, at the filtering step, use the rectangular shape of the MBRs to rapidly obtain the cells of the raster overlapping each MBR, and thus they are also able to produce two lists of definitive and probable results. Since these lists must be, necessarily, the same as those produced by our algorithm, all our measurements in this experiment have excluded, for practical reasons, the effects of the final refinement step.

\subsubsection*{Memory usage}

In this experiment, we do not include the values of Brisaboa et al. \cite{RodriguezBrisaboa17} since the authors' implementation of the $k^2$-acc is very inefficient in managing the memory consumption, yielding really bad results.

\begin{figure}[!h]
	\centering
		\includegraphics[width=1\textwidth]{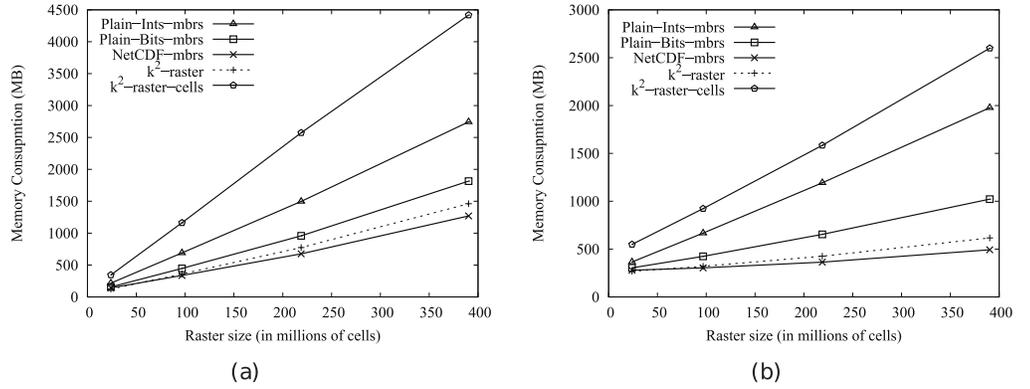}
	\caption{{\bf Memory consumption (in Megabytes) for rasters in Scenario I.}
	(a) \texttt{vects} dataset and (b) \texttt{vecca} dataset.}
	\label{fig:mem_I}
\end{figure}

Fig \ref{fig:mem_I} shows the main memory consumption for Scenario I. Our framework is denoted as \texttt{k$^2$-raster}. We can observe that, compared to the baselines with uncompressed representations, our approach gets always the lowest memory consumption and the best scalability when joining our raster collections with both vector datasets.

However, comparatively speaking, the improvement in memory consumption is lower than the one observed when we compared disk space consumptions. This could be partially explained by the fact that the reported memory consumption includes the size of the output. As such size, which is the same for all approaches, is usually large, it represents a good part of the memory space consumed by all of them. This means that, when expressing percentage differences between the approaches, the distances are reduced, distorting the comparison.

Nevertheless,  our method uses only 31--73\% of the memory space used by \texttt{Plain-Ints-mbrs}, and 60--89\% of that used by \texttt{Plain-Bits-mbrs}. This is clearly an important improvement, and shows, in addition, that the memory consumption of \texttt{k$^2$-raster} is not seriously harmed by the fact of having to manage data in a compressed format.

The situation changes in the comparison  with \texttt{NetCDF}. When checking for portions of the raster, \texttt{NetCDF} loads the needed blocks of the file one by one. Each block is decompressed, processed, and then removed from main memory. Therefore, this baseline consumes less memory in the largest datasets, although  differences are below 25\%.

Finally, \texttt{k$^2$-raster-cells} has the worst behaviour, since  \texttt{k$^2$-raster} consumes between 23\% and 48\% of the space required by the baseline.

\begin{figure}[!h]
	\centering
		\includegraphics[width=1\textwidth]{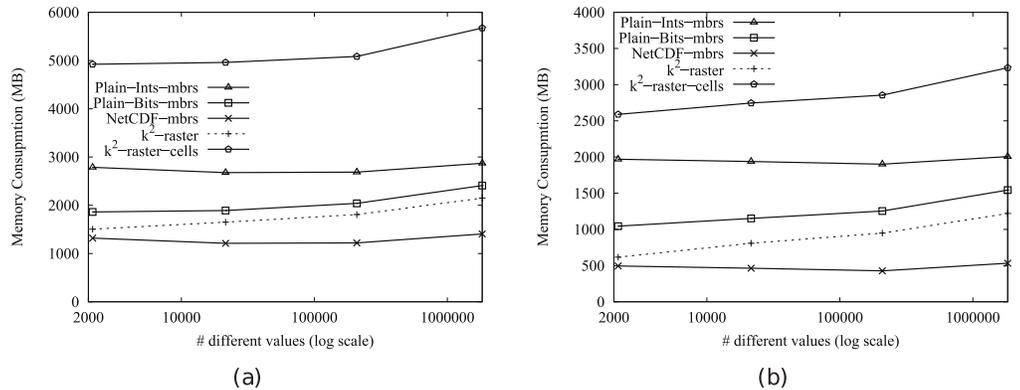}
	\caption{{\bf Memory consumption (in Megabytes) for rasters in Scenario II.} 
	(a) \texttt{vects} dataset and (b) \texttt{vecca} dataset.}
	\label{fig:mem_II}
\end{figure}

Fig \ref{fig:mem_II} shows the results for Scenario II. Our algorithm consumes  between 31\% and 89\% of the space used by the uncompressed baselines. On the contrary, \texttt{NetCDF-mbrs} consumes between 43\% and 87\% of the space required by \texttt{k$^2$-raster}.

Again, the worst values are those of the strategy \texttt{cells} using the $k^2$-raster, since our method uses between 24\% and 38\% of the space consumed by \texttt{k$^2$-raster-cells}.

\subsubsection*{Time performance}

In order to obtain compression, the $k^2$-raster needs a complex arrangement of the data it stores. Therefore, our uncompressed baselines are tough competitors since they keep the raster data in main memory uncompressed and arranged as a simple space-filling curve  row by row.

This problem is common to all compact data structures. However,  many of them obtain better usage times than managing directly uncompressed data. The main reasons are: (i) the dataset representation is smaller, yielding a better usage of the memory hierarchy; and (ii) most of them include, in the same compressed space, indexes that speed up queries.

In our case, $k^2$-raster uses the quadtree arrangement of the data, which cleverly obtains compression and, at the same time, obtains a spatial index. In addition, it couples the quadtree with lightweight indexation, to index the values at cells. Therefore, part of the improvements in processing time of our framework comes from the capacity of the indexes of the $k^2$-raster to prune the search space.

\begin{figure}[!h]
	\centering
		\includegraphics[width=1\textwidth]{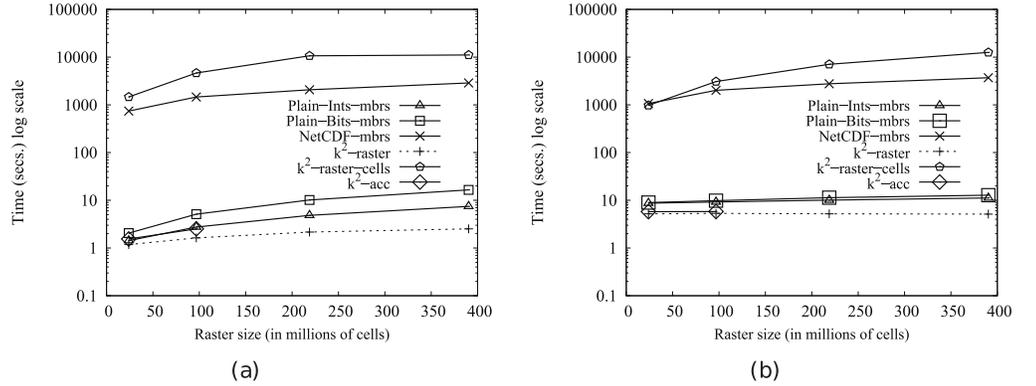}   
	\caption{{\bf Processing time (in seconds and log scale)   with rasters of Scenario I.}
	(a) \texttt{vects} dataset and (b) \texttt{vecca} dataset.}
	\label{fig:time_w_I}
\end{figure}

\begin{figure}[!h]
	\centering
		\includegraphics[width=1\textwidth]{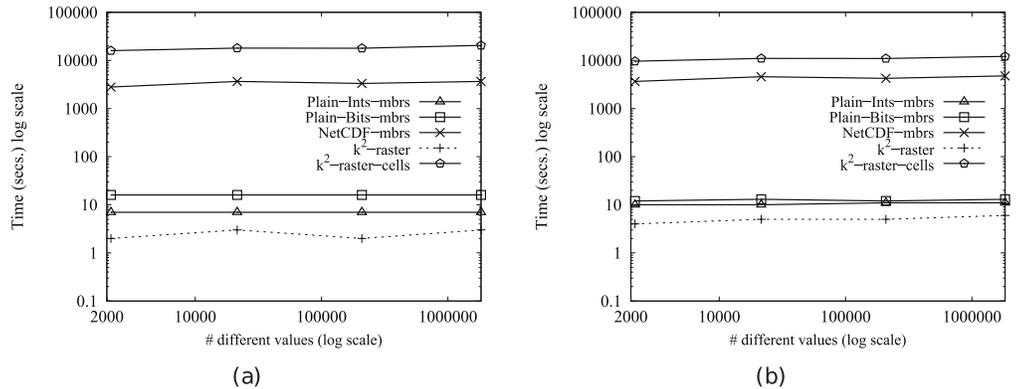} 
	\caption{{\bf Processing time (in seconds and log scale)   with rasters of Scenario II.}
	(a) \texttt{vects} dataset and (b) \texttt{vecca} dataset.}
	\label{fig:time_w_II}
\end{figure}

 Figs \ref{fig:time_w_I} and \ref{fig:time_w_II} show outstanding improvements, even using  a logarithmic scale on the y-axis. The indexes of the $k^2$-raster do their job and thus our method is between 1.21 and 3.55 times faster than \texttt{Plain-Ints-mbrs} and between 1.72 and 8 times faster than \texttt{Plain-Bits-mbrs}.

\texttt{NetCDF}, which uses a traditional compression method, competes with our framework in space consumption, both in disk and main memory, obtaining better results in some cases. However, when measuring processing time, results are clearly worse.
As explained, each time a portion of the raster is demanded, this requires to  decompress one or more blocks storing that information. This costly process should be carried out for each MBR in the leaves of the R-tree. The blocks must also be loaded from disk, but remember that disk access times are not taken into account in this experiment. As we can see in the plots, our approach outperforms this baseline in around three orders of magnitude. The results show that a reasonable sacrifice of memory space clearly improves response times. 

The \texttt{cells} baseline using $k^2$-raster has the worst behaviour, even one order of magnitude worse than \texttt{NetCDF}. 

Finally, the approach of Brisaboa et al., denoted in the experiments as \texttt{$k^2$-acc}, was only able to run over the two smallest datasets of Scenario I, as it works really badly already when the raster has a moderate number of different values \cite{RodriguezBrisaboa17, k2rasterIS}.

\subsubsection*{Main memory consumption versus processing time trade-off}

\begin{figure}[!h]
	\centering
		\includegraphics[width=1\textwidth]{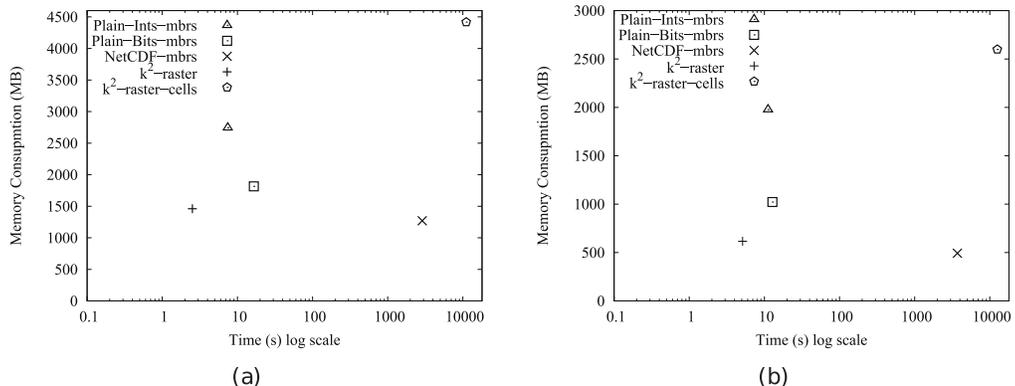}
	\caption{{\bf Memory consumption vs Processing time  with the largest raster of Scenario I.}
	Memory consumption (in Megabytes) vs Processing time (in seconds and log scale)  with the largest raster of Scenario I.
	(a) \texttt{vects} dataset and (b) \texttt{vecca} dataset.}
	\label{fig:TimeVsMemory_0}
\end{figure}

As a summary, Fig \ref{fig:TimeVsMemory_0} includes two plots showing the trade-off between main memory consumption and processing time achieved with the largest dataset of Scenario I, using logarithmic scale in the x-axis. Again the $k^2$-acc is omitted since its memory consumption is very high.
Clearly, we can see that our method is the best choice.

\subsubsection*{Cold start}

Previous experiments were run in ``warm'' start, that is, the queries were run sequentially, and thus, the operating system disk buffer keeps, with high probability, important parts of the input datasets in main memory. This eliminates the effect of disk access on times.

In order to show the impact of disk access, we designed an experiment in ``cold'' start. This implies that:

\begin{itemize}
    \item We measured {\sc real} time to  include disk access times.
    \item All data structures resided on disk and the operating system buffer was cleared before the execution of each query.
\end{itemize}

We run 4 queries over each collection (recall that each collection contains several datasets).

\begin{figure}[!h]
	\centering
		\includegraphics[width=1\textwidth]{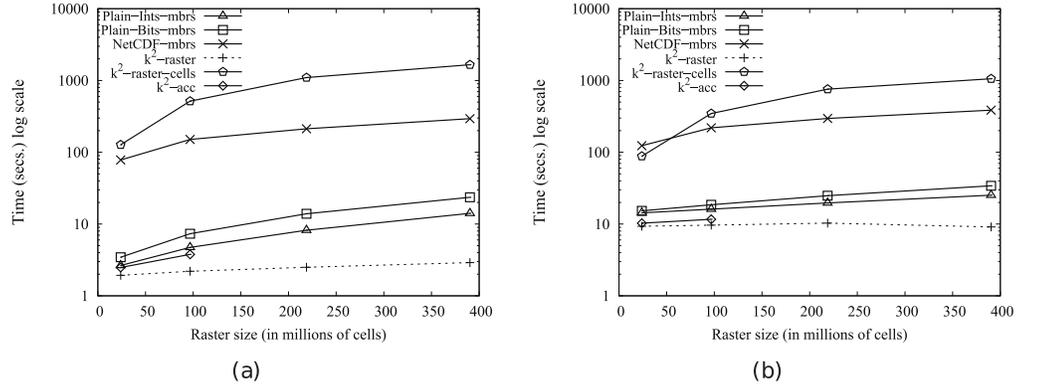}
	\caption{{\bf Processing time (in seconds and log scale)   with rasters of Scenario I and cold start.}
	(a) \texttt{vects} dataset and (b) \texttt{vecca} dataset.}
	\label{fig:time_w_I_cold}
\end{figure}

Fig \ref{fig:time_w_I_cold} shows the results, which do not change the general picture with respect to the performance of each method shown in  Fig \ref{fig:time_w_I}.

\subsubsection*{Comparison with classical approaches}

One of the closest works is that of Corral et al. \cite{corral1999algorithms}, but it only considers binary rasters. Even thought the comparison can be considered unfair for our framework, which can also represent integer rasters, we ran some experiments to study how our approach performs compared to that baseline. 

To obtain binary rasters, we took the datasets of our Scenario I and set to zero the cells with values below the median, and to one the others. Table \ref{Tab:binnary} shows the size of the resulting datasets. The work of Corral et al.\ uses a classical setup, including a linear quadtree stored in a $B^+$-tree to represent the binary raster. As a modern compact data structure, the $k^2$-raster obtains important savings in space.

\begin{table}[h]
\caption{Size of the datasets in Kilobytes.}\label{Tab:binnary}
\centering
\begin{tabular}{|l|r|r|}
\hline
 & Linear Quadtree & $k^2$-raster \\
 \hline
Binary	\texttt{DTM-1$\times$1} & 3595.1           & 39.3       \\
\hline
 Binary	\texttt{DTM-2$\times$2} & 12170.5           & 121.3      \\
 \hline
 Binary	\texttt{DTM-3$\times$3}& 24076.1          & 215.0      \\
 \hline
 Binary	\texttt{DTM-4$\times$4}& 27340.3           & 244.7     \\
 \hline
\end{tabular}
\end{table}

\begin{figure}[!h]
	\centering
		\includegraphics[width=1\textwidth]{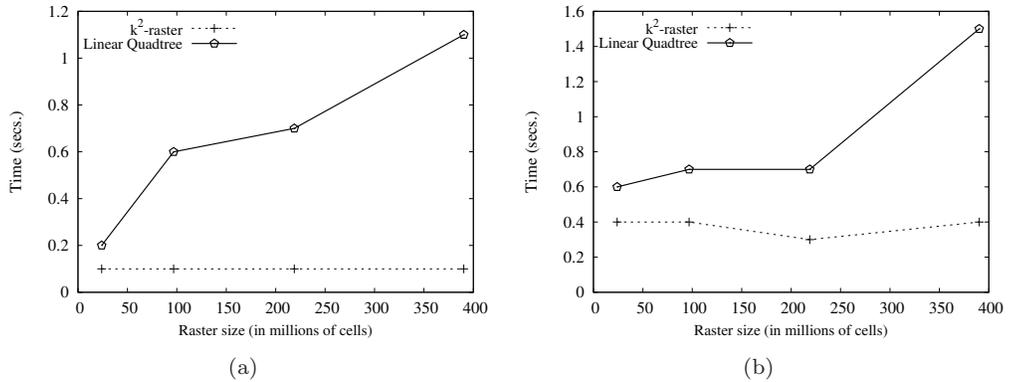} 
	\caption{{\bf Processing time (in seconds)  of the comparison with a classical data structure approach   in Scenario I.}
	(a) \texttt{vects} dataset and (b) \texttt{vecca} dataset.}
	\label{fig:time_w_I_bin}
\end{figure}

As seen in Fig \ref{fig:time_w_I_bin}, our method is faster even though that the $k^2$-raster has a complex data structure  designed to compress integers. Observe that each access to a node requires two subtractions to obtain the values at that node, and in the leaves, it also requires accessing to a compressed representation for integer sequences (DACs), which is also costly. For binary rasters, instead of a $k^2$-raster, we could use a simpler and faster $k^2$-tree, which does not need neither subtractions nor DACs representation.

However, the lightweight index at the nodes of the $k^2$-raster plus its small size  are still able to improve times.

\subsection*{Top-$K$}

All the algorithms have been implemented, for practical reasons, in a simplified way.  We considered that the vector objects are the MBRs of the leafs of the R-tree. 
With this, we avoid running the computational geometry algorithm, whose implementation is the same in the code of both the baselines and \texttt{k$^2$-raster}. Therefore, the $checkGeometry$ function uses a simple map between the MBRs and the corresponding overlapping cells.

\subsubsection*{Memory usage}

\begin{figure}[!h]
	\centering
		\includegraphics[width=1\textwidth]{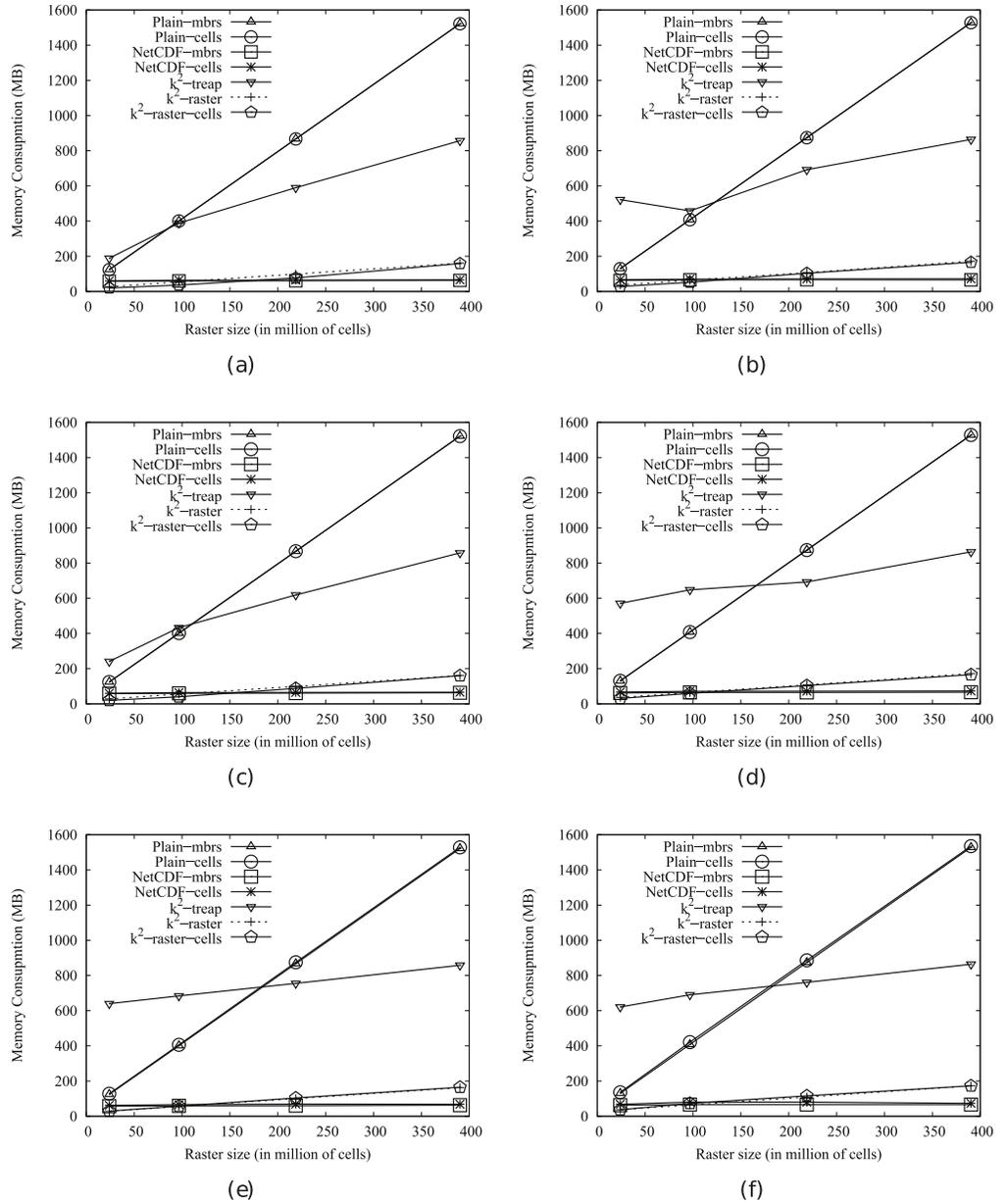}
	\caption{{\bf Average memory consumption for retrieving the top-$K$ over collections of Scenario I.}
	Average memory consumption (in Megabytes) for retrieving the top 1, 10 and 100 MBRs over collections of Scenario I.
	(a) top-1 and \texttt{vects} dataset, (b) top-1 and \texttt{vecca} dataset,
	(c) top-10 and \texttt{vects} dataset, (d) top-10 and \texttt{vecca} dataset,
	(e) top-100 and \texttt{vects} dataset  and (f) top-100 and \texttt{vecca} dataset.}
	\label{fig:memory}
\end{figure}

Fig \ref{fig:memory} shows the memory consumption for top-1, top-10, and top-100 for the datasets of  Scenario I. The uncompressed baselines that completely store the raster in main memory are the worse alternatives, using between 3 and 9 times more memory than our framework. Although \texttt{k$^2$-treap} performs worse than the uncompressed baselines dealing with small files, it scales better, but still,  it is  between 5 and 15 times worse than \texttt{k$^2$-raster}.
	
The traditional  processing of NetCDF files by blocks makes this alternative the best one dealing with large files. With small files, \texttt{k$^2$-raster} consumes around 2 times less, whereas in large files the behavior is just the opposite. 
Finally, the \texttt{cells} strategy with $k^2$-raster has exactly the same behaviour as our method.

\begin{figure}[!h]
	\centering
		\includegraphics[width=1\textwidth]{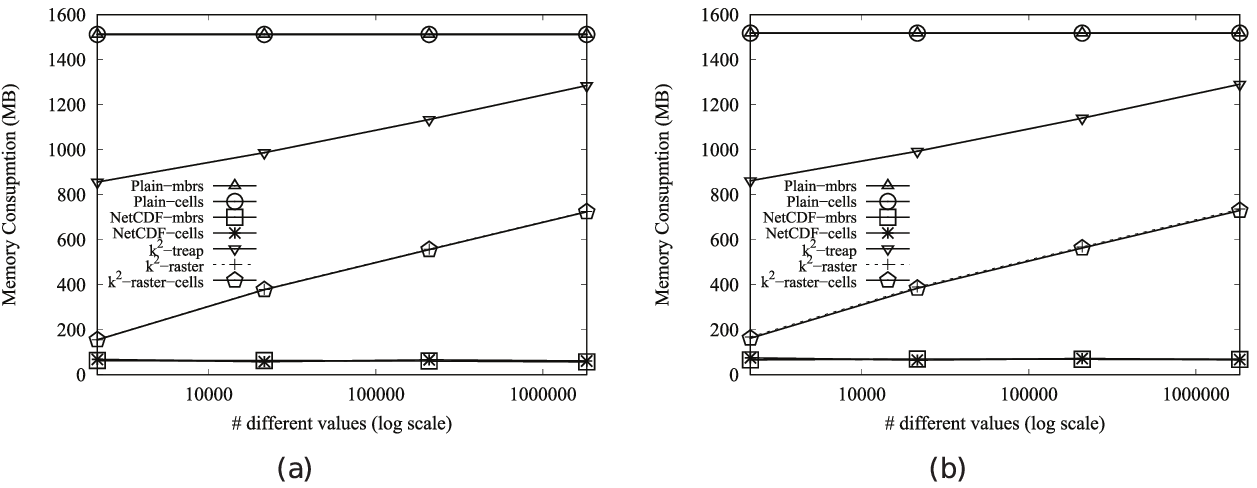}
	\caption{{\bf Average memory consumption for retrieving the top-$K$ over collections of Scenario II.}
	Average memory consumption (in Megabytes) for retrieving the top  10  MBRs over collections of Scenario II. 
	(a) \texttt{vects} dataset and (b) \texttt{vecca} dataset.}
	\label{fig:memoryMDT}
\end{figure}

Fig \ref{fig:memoryMDT} shows the memory consumption for the datasets of Scenario II. We only include the results for top-10. In this experiment, NetCDF is the clear winner, consuming around 7 times less than \texttt{k$^2$-raster}.

\subsubsection*{Time performance}
\begin{figure}[!h]
    \centering    
		\includegraphics[width=1\textwidth]{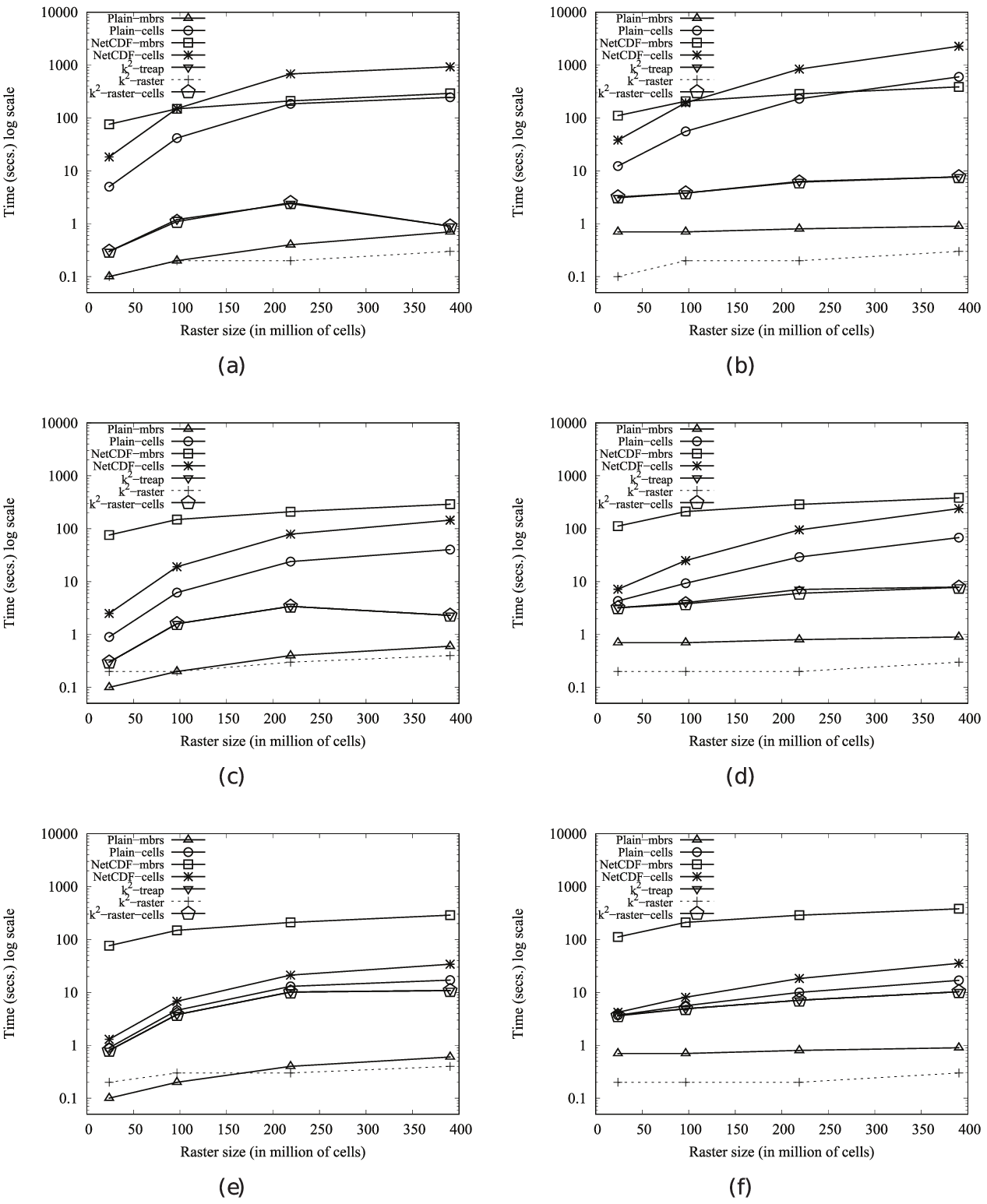}
    \caption{{\bf Average time results for retrieving the top-$K$ over collections of Scenario I.}
    Average time results (in seconds) for retrieving the top 1, 10 and 100 MBRs over collections of Scenario I. We compare the results for all the algorithms using logarithmic scales for all the figures.
	(a) top-1 and \texttt{vects} dataset, (b) top-1 and \texttt{vecca} dataset,
	(c) top-10 and \texttt{vects} dataset, (d) top-10 and \texttt{vecca} dataset,
	(e) top-100 and \texttt{vects} dataset  and (f) top-100 and \texttt{vecca} dataset.}
\label{fig:times}
\end{figure}

Fig \ref{fig:times} shows the average time results for our experiments, performing, respectively, top-1, top-10 and top-100 operations. For clarity, times are shown in logarithmic scale. 

The main advantage of our algorithm is that it uses two indexes simultaneously and synchronously. During a filtering process, discarding nodes of one index allows us to also discard some nodes of the other, and the other way around. This fact saves time with respect to the naive strategies of the baselines, which take as  starting point one of the datasets, and process some of their elements (MBRs, cells) in a certain order. Each element examined implies a search in the other dataset (with or without the help of an index), sometimes unsuccessfully. This means a great loss of time. This conclusion is confirmed by the experiments: our algorithm obtains the best top-$K$ retrieval times in all scenarios, up to four orders of magnitude better than \texttt{NetCDF-cells} and \texttt{Plain-cells}, up to 5 times faster than \texttt{Plain-mbrs}, and up to 33 times than \texttt{k$^2$-treap} and \texttt{k$^2$-raster-cells}. Some other facts support these results:

\begin{itemize}
\item [--] First of all, \texttt{mbrs} strategy works better than \texttt{cells}, as it can be seen with \texttt{Plain-mbrs} and \texttt{Plain-cells}. Finding the MBRs that overlap with each cell is fast thanks to the R-tree; but it may happen that most of the cells with the highest values do not overlap with any leaf MBR, so that many searches in the R-tree finally become a waste of time. Instead, \texttt{Plain-mbrs} starts processing leaf MBRs (and there are fewer MBRs than cells), and for each of those MBRs it must check just a few raster cells (so not using an index does not penalize too much). In fact, it is close to the \texttt{k$^2$-raster} performance, behaving worse mainly because it never filters leaf MBRs, and thus it must check raster cells for each one of them. 

\item [--] Second, compression penalizes the baselines \texttt{NetCDF-mbrs} and \texttt{NetCDF-cells}, which always perform worse than their alternative plain versions, that is, \texttt{Plain-mbrs} and \texttt{Plain-cells}. However, \texttt{k$^2$-raster} offers close compression rates that do not penalize search times.

\item [--] Finally, as expected, although \texttt{Plain-cells}, \texttt{k$^2$-raster-cells}, and \texttt{k$^2$-treap} follow the same strategy, the compressed data structures  behave better, since they use an additional index that allows accessing the raster cells in an orderly and cost-free manner. However, those indexes are not enough to reach the performance of \texttt{k$^2$-raster} and its synchronized filtering process.
\end{itemize}

\begin{figure}[!h]
	\centering   
		\includegraphics[width=1\textwidth]{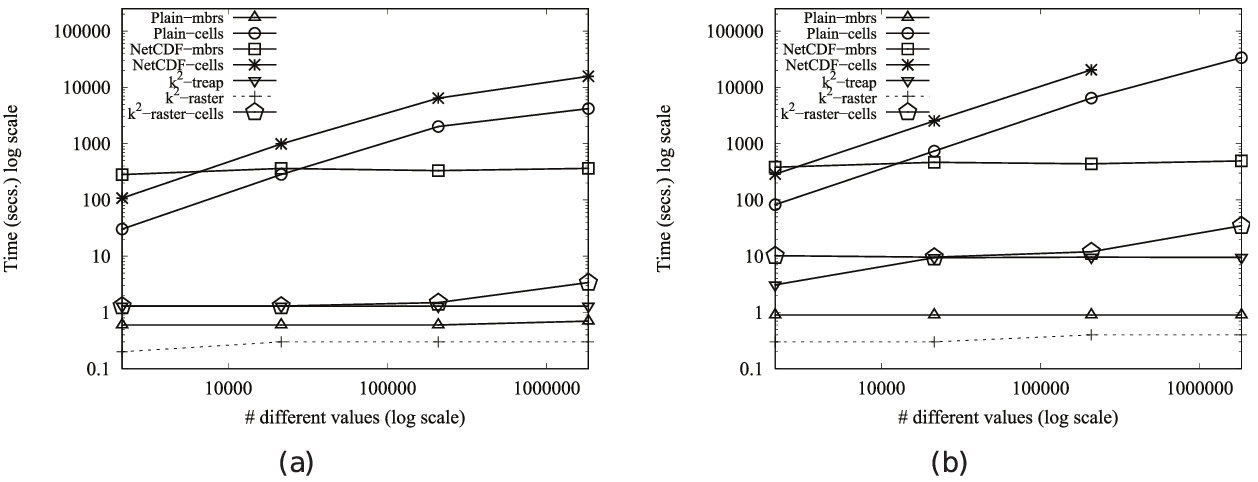}
	\caption{{\bf Time performance for retrieving the top-$K$  MBRs over collections of Scenario II.}
	Time performance (in seconds) for retrieving the top  10  MBRs over collections of Scenario II. Both axes uses a logarithmic scale. 
	(a) \texttt{vects} dataset and (b) \texttt{vecca} dataset.}
	\label{fig:timeMDT}
\end{figure}

Fig \ref{fig:timeMDT} shows the times for the datasets of Scenario II, for top-10. The value of \texttt{NetCDF-cells} for the \texttt{vecca} raster with largest number of different values is not displayed because it lasted too long.\\

\begin{figure}[!h]
    \centering     
		\includegraphics[width=1\textwidth]{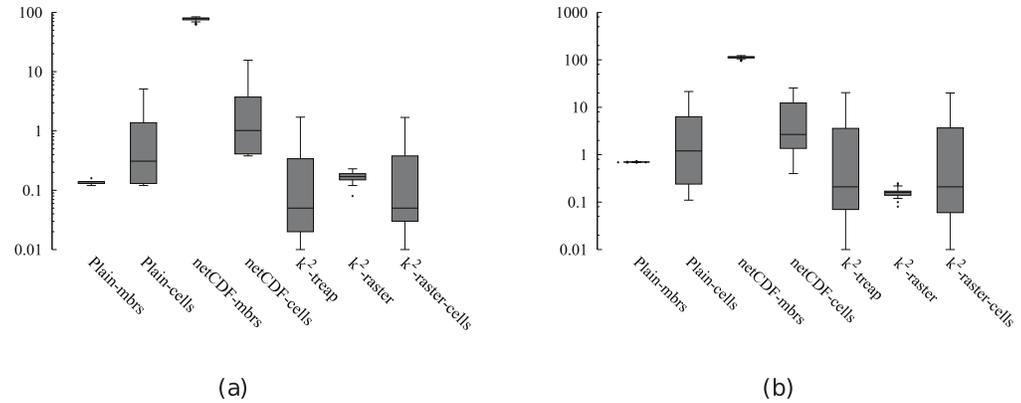}
    \caption{{\bf Box plots showing time results for retrieving  the top-10  MBRs.} 
    Box plots showing time results (in seconds) for retrieving  the top-10  MBRs for the 25 matrices of collection \texttt{DTM-1$\times$1}. The y axis is in logarithmic scale. 
	(a) \texttt{vects} dataset and (b) \texttt{vecca} dataset.}
    \label{fig:timesbox}
\end{figure}

Fig \ref{fig:timesbox}  shows the box plots for executing the top-$10$ queries over each raster matrix of the \texttt{DTM-1$\times$1} of Scenario I. Baselines using \texttt{cells} strategy obtain hugely irregular query times, as their performance highly depends on the existence of leaf MBRs overlapping the cells first returned by the algorithm (those having the largest values). \texttt{k$^2$-treap} and \texttt{k$^2$-raster-cells}  have approximately the  same results, since both, as explained, use a \texttt{cells} strategy with the help of an index which allows them to access the cells in an orderly manner. They show a much worse behaviour on average than \texttt{k$^2$-raster}, although they perform better  in certain occasions, specially when dealing with the \texttt{vects} dataset. When  most top-$K$ cells match leaf MBRs, \texttt{k$^2$-treap}  and \texttt{k$^2$-raster-cells} are always
  faster than \texttt{k$^2$-raster} solving queries. However, our  synchronous algorithm promotes this best-case scenario, discarding ``bad'' cells beforehand, and making \texttt{k$^2$-raster} better on average. This suggests that it could be promising to adapt our algorithm to \texttt{k$^2$-treap} too. Unfortunately, \texttt{k$^2$-treap} is not a feasible option for the framework, since it would only be useful in this query, being inefficient for the top-$K$ version with the minimum values, or  the join query, among others.
  
\subsubsection*{Main memory consumption versus processing time trade-off}

\begin{figure}[!h]
	\centering   
		\includegraphics[width=1\textwidth]{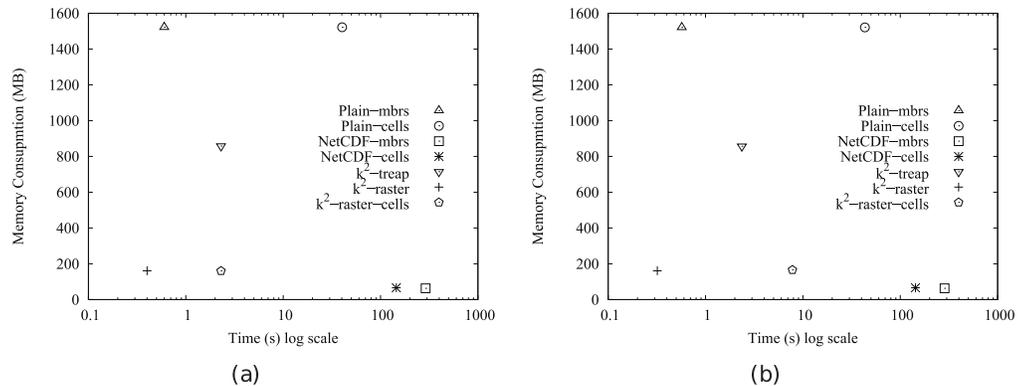}
	\caption{{\bf Memory consumption vs Processing time with the largest raster of Scenario I and top-10.}
	Memory consumption (in Megabytes) vs Processing time (in seconds)  with the largest raster of Scenario I and top-10. 
	(a) \texttt{vects} dataset and (b) \texttt{vecca} dataset.}
	\label{fig:TimeVsMemory_0_top}
\end{figure}

Finally, Fig \ref{fig:TimeVsMemory_0_top} includes two plots showing the trade-off between main memory consumption and processing time achieved with the largest dataset of Scenario I and top-10, using logarithmic scale in the \textit{x} axis.
As seen,  our method method is by far the best choice.

\subsubsection*{Cold start}

\begin{figure}[!h]
	\centering 
		\includegraphics[width=1\textwidth]{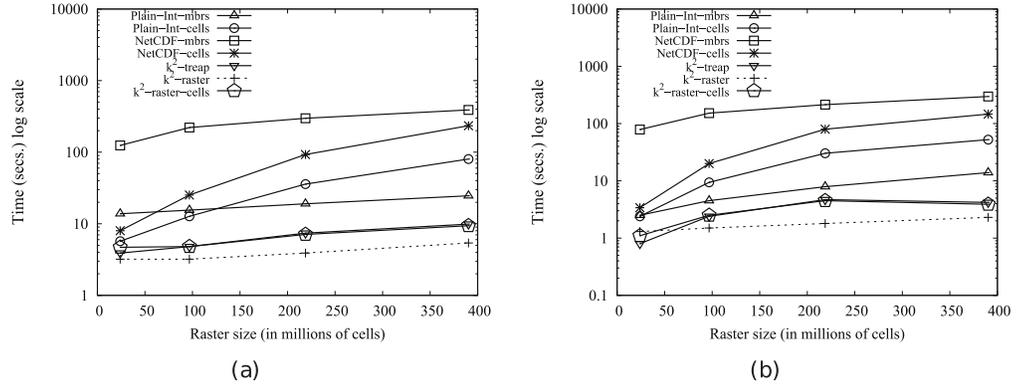}
	\caption{{\bf Time performance for retrieving the top-10  MBRs over collections of Scenario I in cold start.}
 The y axis uses a logarithmic scale. 
	(a) \texttt{vects} dataset and (b) \texttt{vecca} dataset.}
	\label{fig:TimeVsMemory_0_top_cold}
\end{figure}

Fig \ref{fig:TimeVsMemory_0_top_cold} shows the results of the experiment in ``cold'' start for top-10. In this experiment, the differences between our method and $k^2$-treap are shortened. 

\section*{Results discussion} \label{sec:Discussion}

In  our experiments with the join operation,  our new framework  consumes   between 31--89\% of memory used by our \texttt{Plain} baselines, for which, in addition,  the  time performance  is  improved up to 8 times.

In the case of \texttt{NetCDF}, our framework is around three orders of magnitude faster, while it only implies a slight sacrifice in  space. 

In the case of top-$K$, our framework   is up to five orders of magnitude faster than the  \texttt{Plain-cells} and \texttt{NetCDF-cells}, and up to 33 times faster than the \texttt{k$^2$-treap} and \texttt{k$^2$-raster-cells}. 

The memory consumption shows important improvements; our framework  requires between 3 and 9 times less memory than  the  \texttt{Plain} baselines, in the case of the top-$K$ operation, and up to 3 times in the join. In the case of the \texttt{k$^2$-treap}, used in the top-$K$ operation, our method uses between 5 and 15 times less space, which shows that using a compact data structure does not directly imply a gain in memory consumption, and thus a careful design of query algorithms is required.

The reason of these improvements is due to the fact that our algorithms take advantage of a smaller input, and thus the memory hierarchy between main memory and the processor is more efficient; and that they make intelligent use of the indexes of the $k^2$-raster, which includes, in the same compressed space,  a spatial index and an index over the values of the raster.

\section*{Conclusions} \label{sec:Conclusions}

The possibility of managing raster and vector datasets in geographical information systems is a convenient feature, since it is well-known that each model is more adequate depending on the nature of the spatial data \cite{Couclelis92}. However, commercial and open-source systems, and even the OGC standard \cite{OGCWFS,OGCWCS}, separate both views and do not provide languages, data structures, and algorithms to perform queries that use information from both models.

The exception to this rule could be the zonal statistics operation of Map Algebra \cite{tomlin1994map} that is included in several systems. However, those systems internally translate the vector dataset into a raster dataset before running the operation.
In this work, we have presented a framework that includes two known data structures and two new algorithms for running a join between vector and raster datasets, and  for retrieving \textit{K} objects of a vector dataset that overlap  cells  of a raster dataset, such that the $K$ objects are those overlapping the highest (or lowest) cell values among all objects, without a previous transformation of any of them. The good properties shown by this new approach are due to the use of compact data structures, which allows efficient processing in little space.

The idea of compact data structures, like the $k^2$-raster, is to keep the information always compressed, even when processing it. The $k^2$-raster have been proven to be a good approach when performing operations involving the raster alone \cite{SSDBM16,k2rasterIS}. However, in this work, we show that the use of $k^2$-rasters for storing raster data brings new opportunities, in this case, efficient operations between a raster and a vector dataset without translating none of them.

The use of the $k^2$-raster represented a challenge. It was designed to be used in main memory without a previous decompression of the whole dataset, but this requires a complex arrangement of the data (something analogous to most compact data structures). This means that the decompression processes applied on small parts upon demand, and the management of the indexes it contains, can degrade the main memory consumption, and affect the processing times. If we did not design the  algorithms carefully, both problems could arise. As shown in our experiments, we have successfully solved it.

The election of the R-tree for indexing the vector dataset is a pragmatic choice, since it is the \textit{de facto} standard for this type of data. However, as future work we will consider the use of modern compact data structures as a substitution for the R-tree.
We also plan to add new operations between raster and vector datasets to our framework.

\section*{Acknowledgments}
The authors wish to thank Antonio Corral for providing the code of their spatial join.

\section*{Funding}
This work has received funding from the European Union’s Horizon 2020 research and innovation programme under the Marie Skłodowska-Curie grant agreement No 690941; from the Ministerio de Ciencia, Innovación y Universidades (PGE and ERDF) grant numbers TIN2016-78011-C4-1-R; TIN2016-77158 C4-3-R; RTC-2017-5908-7; from Xunta de Galicia (co-founded with ERDF) grant numbers ED431C 2017/58; ED431G/01; IN852A 2018/14; and University of Bío-Bío grant numbers 192119 2/R; 195119 GI/VC.


\bibliographystyle{plos2015}

\begin{thebibliography}{10}

\bibitem{Couclelis92}
Couclelis H.
\newblock People Manipulate Objects (but Cultivate Fields): Beyond the
  Raster-Vector Debate in {GIS}.
\newblock In: International Conference GIS: from space to territory - theories
  and methods of spatio-temporal reasoning. London: Springer-Verlag; 1992. p.
  65--77.

\bibitem{bb78851}
Li Y, Bretschneider TR.
\newblock {Semantic-Sensitive Satellite Image Retrieval}.
\newblock IEEE Transactions on Geoscience and Remote Sensing.
  2007;45(4):853--860.

\bibitem{quartulli2013review}
Quartulli M, {G  Olaizola} I.
\newblock {A review of EO image information mining}.
\newblock ISPRS Journal of Photogrammetry and Remote Sensing. 2013;75:11--28.

\bibitem{grumbach2000manipulating}
Grumbach S, Rigaux P, Segoufin L.
\newblock Manipulating interpolated data is easier than you thought.
\newblock In: 26th International Conference on Very Large Data Bases. San
  Francisco, CA: Morgan Kaufmann Publishers Inc.; 2000. p. 156--165.

\bibitem{zonalS}
{Environmental Systems Research Institute Inc }. Zonal Statistics Help|ArcGis
  for Desktop; 2016.
\newblock Available from:
  \url{http://webhelp.esri.com/arcgisdesktop/9.3/index.cfm?TopicName=Zonal\_Statistics}.

\bibitem{zonalSG}
{GRASS development team}. GRASS GIS manual:v.rast.stats; 2016.
\newblock Available from:
  \url{https://grass.osgeo.org/grass72/manuals/v.rast.stats.html}.

\bibitem{corral1999algorithms}
Corral A, Vassilakopoulos M, Manolopoulos Y.
\newblock Algorithms for joining R-trees and linear region quadtrees.
\newblock In: 6th International Symposium on Advances in Spatial Databases.
  London: Springer-Verlag; 1999. p. 251--269.

\bibitem{TPRtrees}
Corral A, Torres M, Vassilakopoulos M, Manolopoulos Y.
\newblock Predictive Join Processing between Regions and Moving Objects.
\newblock In: 12th East European conference on Advances in Databases and
  Information Systems. Berlin, Heidelberg, Germany: Springer-Verlag; 2008. p.
  46--61.

\bibitem{RodriguezBrisaboa17}
Brisaboa NR, de~Bernardo G, Guti\'errez G, Luaces MR, Param\'a JR.
\newblock Efficiently Querying Vector and Raster Data.
\newblock The Computer Journal. 2017;60(9):1395--1413.

\bibitem{Eldawy:2017}
Eldawy A, Niu L, Haynes D, Su Z.
\newblock Large Scale Analytics of Vector+Raster Big Spatial Data.
\newblock In: Proceedings of the 25th ACM SIGSPATIAL International Conference
  on Advances in Geographic Information Systems. SIGSPATIAL'17; 2017. p.
  62:1--62:4.

\bibitem{plattner2012memory}
Plattner H, Zeier A.
\newblock In-Memory Data Management: Technology and Applications.
\newblock Heidelberg, Germany: Springer-Verlag Berlin; 2012.

\bibitem{k2ones}
de~Bernardo G, {\'A}lvarez-Garc\'{\i}a S, Brisaboa NR, Navarro G, Pedreira O.
\newblock Compact Querieable Representations of Raster Data.
\newblock In: 20th International Symposium on String Processing and Information
  Retrieval. New York, NY: Springer-Verlag; 2013. p. 96--108.

\bibitem{SSDBM16}
Ladra S, Param{\'{a}} JR, Silva-Coira F.
\newblock Compact and queryable representation of raster datasets.
\newblock In: 28th International Conference on Scientific and Statistical
  Database Management. New York, NY, USA: ACM; 2016. p. 15:1--15:12.

\bibitem{k2rasterIS}
Ladra S, Param{\'{a}} JR, Silva-Coira F.
\newblock Scalable and Queryable Compressed Storage Structure for Raster Data.
\newblock Information Systems. 2017;72:179--204.

\bibitem{pinto2017improved}
Pinto A, Seco D, Guti{\'e}rrez G.
\newblock Improved Queryable Representations of Rasters.
\newblock In: Data Compression Conference. IEEE. IEEE; 2017. p. 320--329.

\bibitem{Nav16}
Navarro G.
\newblock Compact Data Structures -- A practical approach.
\newblock New York, NY: Cambridge University Press; 2016.

\bibitem{plattner2013course}
Plattner H.
\newblock A Course in In-Memory Data Management: The Inner Mechanics of
  In-Memory Databases.
\newblock Heidelberg, Germany: Springer-Verlag Berlin; 2013.

\bibitem{Guttman:1984:RDI:602259.602266}
Guttman A.
\newblock R-trees: A Dynamic Index Structure for Spatial Searching.
\newblock In: 1984 ACM SIGMOD international conference on Management of data.
  New York, NY: ACM; 1984. p. 47--57.

\bibitem{Manolopoulos:2005:RTA:1098699}
Manolopoulos Y, Nanopoulos A, Papadopoulos AN, Theodoridis Y.
\newblock R-trees: Theory and Applications (Advanced Information and Knowledge
  Processing).
\newblock Secaucus, NJ: Springer-Verlag New York; 2005.

\bibitem{worboys2004gis}
Worboys MF, Duckham M.
\newblock GIS: a computing perspective.
\newblock Boca Raton, FL: CRC press; 2004.

\bibitem{ISO19107}
{ISO}.
\newblock Geographic information -- Spatial schema.
\newblock Geneva, Switzerland; 2003. ISO 19107:2003.

\bibitem{ISO19123}
{ISO}.
\newblock Geographic information -- Schema for coverage geometry and functions.
\newblock Geneva, Switzerland; 2005. ISO 19123:2005.

\bibitem{OGCWFS}
OGC.
\newblock {OpenGIS Web Feature Service 2.0 Interface Standard}.
\newblock Wayland, MA; 2010. OGC 09-025r2.

\bibitem{OGCWCS}
OGC.
\newblock {OpenGIS Web Coverage Service 2.0 Interface Standard - Core:
  Corrigendum}.
\newblock Wayland, MA; 2012. OGC 09-110r4.

\bibitem{tomlin1979mathematical}
Tomlin CD, Berry JK.
\newblock Mathematical structure for cartographic modeling in environmental
  analysis.
\newblock In: the American Congress on Surveying and Mapping 39th Annual
  Meeting. ACSM; 1979. p. 269--283.

\bibitem{tomlin90}
Tomlin DC.
\newblock Geographic Information Systems and Cartographic Modeling.
\newblock Englewood Cliffs, NJ: Prentice-Hall; 1990.

\bibitem{svensson1991geo}
Svensson P, Zhexue H.
\newblock Geo--{SAL:} {A} Query Language for Spatial Data Analysis.
\newblock In: SSD 1991. vol. 525 of LNCS. Berlin, Heidelberg: Springer; 1991.
  p. 119--142.

\bibitem{baumann1998multidimensional}
Baumann P, Dehmel A, Furtado P, Ritsch R, Widmann N.
\newblock The Multidimensional Database System RasDaMan.
\newblock In: 1998 ACM SIGMOD international conference on Management of data.
  New York, NY: ACM; 1998. p. 575--577.

\bibitem{vaisman2009multidimensional}
Vaisman A, Zim{\'a}nyi E.
\newblock A multidimensional model representing continuous fields in spatial
  data warehouses.
\newblock In: 17th ACM SIGSPATIAL International Conference on Advances in
  Geographic Information Systems. New York: ACM; 2009. p. 168--177.

\bibitem{brown2010overview}
Brown PG.
\newblock Overview of SciDB: large scale array storage, processing and
  analysis.
\newblock In: 2010 ACM SIGMOD International Conference on Management of data.
  New York, NY: ACM press; 2010. p. 963--968.

\bibitem{ktree}
Brisaboa NR, Ladra S, Navarro G.
\newblock Compact Representation of Web Graphs with Extended Functionality.
\newblock Information Systems. 2014;39(1):152--174.

\bibitem{Klinger1971303}
Klinger A.
\newblock Patterns and search statistics.
\newblock In: Symposium Held at The Center for Tomorrow The Ohio State
  University. New York, NY, USA: Academic Press; 1971. p. 303--337.

\bibitem{bb26519}
Klinger A, Dyer CR.
\newblock Experiments on Picture Representation Using Regular Decomposition.
\newblock Computer Graphics Image Processing. 1976;5(1):68--105.

\bibitem{Sam2006}
Samet H.
\newblock {Foundations of Multimensional and Metric Data Structures}.
\newblock San Francisco, CA: Morgan Kaufmann; 2006.

\bibitem{Jac89}
Jacobson G.
\newblock {Space-efficient static trees and graphs}.
\newblock In: 30th Annual Symposium on Foundations of Computer Science.
  Washington, DC, USA: IEEE Computer Society; 1989. p. 549--554.

\bibitem{BLN13}
Brisaboa NR, Ladra S, Navarro G.
\newblock {DACs: Bringing direct access to variable-length codes}.
\newblock Information Processing and Management. 2013;49:392--404.

\bibitem{SS17}
Seidemann M, Seeger B.
\newblock Chronicle{DB}: A High-Performance Event Store.
\newblock In: 20th International Conference on Extending Database Technology;
  2017. p. 144--155.

\bibitem{Moe98}
Moerkotte G.
\newblock Small {M}aterialized {A}ggregates: A Light Weight Index Structure for
  Data Warehousing.
\newblock In: 24rd International Conference on Very Large Data Bases. San
  Francisco, CA: Morgan Kaufmann Publishers Inc.; 1998. p. 476--487.

\bibitem{AA14}
Athanassoulis M, Ailamaki A.
\newblock B{F}-tree: Approximate Tree Indexing.
\newblock In: 40th International Conference on Very Large Databases. vol.~7.
  VLDB Endowment; 2014. p. 1881--1892.

\bibitem{BdKNS16}
Brisaboa NR, de~Bernardo G, Konow R, Navarro G, Seco D.
\newblock Aggregated {2D} range queries on clustered points.
\newblock Information Systems. 2016;60:34--49.

\bibitem{lee2008netcdf}
Lee C, Yang M, Aydt R.
\newblock NetCDF-4 Performance Report.
\newblock Technical report, HDF Group; 2008.

\bibitem{RFC1951}
Deutsch LP. {RFC 1951}: {DEFLATE} Compressed Data Format Specification version
  1.3; 1996.

\bibitem{Rigaux}
Rigaux P, Scholl M, Voisard A.
\newblock Spatial databases: with application to GIS.
\newblock Morgan Kauffman; 2002.

\bibitem{B1993}
Brinkhoff T, Kriegel HP, Seeger B.
\newblock Efficient Processing of Spatial Joins Using R-trees.
\newblock SIGMOD Record. 1993;22(2):237--246.

\bibitem{T2000}
Theodoridis Y, Stefanakis E, Sellis TK.
\newblock Efficient Cost Models for Spatial Queries Using R-Trees.
\newblock IEEE Trans Knowl Data Eng. 2000;12(1):19--32.

\bibitem{Theodoridis:1998:CMJ:645483.656214}
Theodoridis Y, Stefanakis E, Sellis TK.
\newblock Cost Models for Join Queries in Spatial Databases.
\newblock In: Proceedings of the Fourteenth International Conference on Data
  Engineering. ICDE '98. Washington, DC, USA: IEEE Computer Society; 1998. p.
  476--483.


\bibitem{Corral2006}
Corral A, Manolopoulos Y, Theodoridis Y, Vassilakopoulos M.
\newblock Cost Models for Distance Joins Queries Using R-trees.
\newblock Data and Knowledge Engineering. 2006;57(1):1--36.


\bibitem{GGMN05}
Gonz{\'{a}}lez R, Grabowski S, M{\"{a}}kinen V, Navarro G.
\newblock {Practical Implementation of Rank and Select Queries}.
\newblock In: 4th Workshop on Efficient and Experimental Algorithms. vol. 0109.
  Berlin, Heidelberg: Springer-Verlag; 2005. p. 27--38.

\bibitem{tomlin1994map}
Tomlin CD.
\newblock Map algebra: one perspective.
\newblock Landscape and Urban Planning. 1994;30(1-2):3--12.

\end{thebibliography}

\end{document}